\documentclass[aps,pre,twocolumn,showpacs,groupedaddress]{revtex4}
\usepackage{graphicx}
\usepackage{dcolumn}
\usepackage{amsmath}
\usepackage{latexsym}

\begin{document}
\title{Coarse-graining microscopic strains in a harmonic, two-dimensional solid and its implications for elasticity: non-local susceptibilities and non-affine noise.}
\author{K.~Franzrahe}
\email[]{Kerstin.Franzrahe@uni-konstanz.de}
\author{P.~Nielaba}
\affiliation{Fachbereich Physik, Universit\"at Konstanz, Postfach 692,78457 Konstanz, Germany}
\author{S.~Sengupta}
\affiliation{
Centre for Advanced Materials, Indian Association for the Cultivation of Science, Jadavpur, Kolkata 700 032, India, \\
{\rm and}  \\
Advanced Materials Research Unit, 
Satyendra Nath Bose National Centre for Basic Sciences, Block-JD,
Sector-III, Salt Lake, Kolkata 700 098, India
}
\date{\today}
\begin{abstract}
In soft matter systems the local displacement field can be accessed directly by video microscopy enabling one to compute local strain fields 
and hence the elastic moduli in these systems using a coarse-graining procedure. Here, we study this process in detail for a simple 
triangular lattice of particles connected by harmonic springs in two-dimensions. Coarse-graining local strains obtained from 
particle configurations in a Monte Carlo simulation generates non-trivial, non-local strain correlations (susceptibilities), which may be 
understood within a generalized, Landau type elastic Hamiltonian containing up to quartic terms in strain gradients 
(K. Franzrahe {\em et. al.}, Phys. Rev. E {\bf 78}, 026106 (2008)). In order to demonstrate the versatility of 
the analysis of these correlations and to make our calculations directly relevant for experiments on colloidal solids, we systematically study in 
detail various parameters such as the choice of statistical 
ensemble, presence of external pressure and boundary conditions. Crucially, we show that special care needs to be taken for an accurate 
application of our results to actual experiments, where the analyzed area is embedded within a larger system, to which it is mechanically 
coupled. Apart from the smooth, affine strain fields, the coarse-graining procedure also gives rise to a {\em noise} field ($\chi$) made 
up of {\em non-affine} displacements. Several properties of $\chi$ may be rationalized for the harmonic solid using a simple "cell model" calculation.
Furthermore the scaling behavior of the probability distribution of the {\em noise} field ($\chi$) is studied. We find that for any inverse
temperature $\beta$, spring constant $f$, density $\rho$ and coarse-graining
length $\Lambda$ the probability distribution can be obtained from a master curve of the scaling variable ${\mathcal X} = \chi \beta f/\rho
\Lambda^2$. 
\end{abstract}
\maketitle

\section{Introduction}
Soft matter with its structural and elastic properties offers an attractive route to the design of new materials. In particular colloidal 
dispersions attract a lot of interest in this context. Surface chemistry or alterations in the composition of the solvent give an excellent control over the effective interactions in colloidal dispersions \cite{YETHI}.
By definition colloids lie in the range of the visible spectrum. Video microscopy \cite{HABD} is therefore a straightforward means to gain information of the microscopic trajectories of the components of the system under study. Thus microscopic, thermal (or Brownian) fluctuations can be resolved directly in
real space, making colloidal dispersions excellent model systems for the study of fundamental questions of the statistical physics of soft condensed matter. Two dimensional
colloidal dispersions, for example, have been used successfully in studies on melting in two dimensions during the last
decades \cite{STRAND,SEN_KTHNY,BIND_KTHNY,GRUENB_MELT}. 
In this paper we focus on the mechanical properties of such systems and consider, in detail, how they may be obtained from the microscopic particle trajectories.

Within linear elasticity, a solid in two-dimensions is described by eight unknown variables: the three stresses $\sigma_{ij}$, three strains
$\epsilon_{ij}$ and two components of the displacement field $u_{i}$. Appropriately, there are also eight equations, namely the two equilibrium
conditions $\partial \sigma_{ji}/{\partial x_{j}}+f_{i}=0$ (with $f_{i}$, the forces per unit volume within the body), three geometrical equations $\epsilon_{ij}=\left(\partial u_{i}/\partial x_{j}+\partial
u_{j}/\partial x_{i}\right)/2$, and three constitutive equations $\sigma_{ij}=C_{ijkl}\epsilon_{kl}$. This set of equations may be solved for a given
boundary condition in order to extract either the stresses or strains, given the elastic moduli, or the elastic moduli themselves, if the strains are
known for a given stress configuration or vice versa. This manner of obtaining elastic moduli requires us to perturb the system using some external
means e.g. laser tweezers \cite{WILLE}. In contrast to this approach, one may calculate the tensor of elastic constants $C_{ijkl}$ of a
system from fluctuations of the microscopic strains obtained by a coarse-graining procedure. Computing $C_{ijkl}$ in this way requires no 
external forces to be applied which may tend to change the very properties that are being measured \cite{SEN_FLUCM,ZAHN,MARAG}.  
Recently, this procedure was further extended in Ref.\cite{KFRANZ2} to obtain even the non-local elastic susceptibilities.
 
The purpose of the present paper is twofold. Firstly, we present in great detail the analytic background of the coarse-graining procedure for obtaining elastic moduli and 
non-local susceptibilities (or strain-strain correlation functions) used in our earlier work \cite{KFRANZ2}. In order to demonstrate the versatility inherent to the 
analysis of these non-local elastic
susceptibilities, we present systematic studies of a simple two dimensional lattice of particles 
connected by harmonic springs in the current paper. A comparison of the non-local susceptibilities in different statistical ensembles for various boundary conditions, system
sizes and under different external conditions is given. Furthermore relations between the non-local susceptibilities and the elastic constants in systems surrounded 
by an embedding medium are derived. The proper interpretation of the correlation functions in such settings is discussed and visualized by use of the 
static susceptibility sum rule. Thus our aim in the present paper is to demonstrate various approaches in the analysis of the
non-local strain correlation functions and to show how the analysis has to be adapted to the actual experimental situation. 
This study will thus greatly facilitate adoption of such techniques for routine analysis of experimental data at least for soft systems, which 
are close to being harmonic. 

Secondly, apart from the above stated intent to establish a precise procedure for obtaining mechanical properties from microscopic
configuration data, we also aim, to study in some detail fundamental aspects of the coarse-graining procedure itself. For example, an immediate
problem is the presence of 
particle configurations within the coarse-graining volume, 
which are not describable in terms of affine deformations of any reference lattice, e.g. incipient vacancy-interstitial pairs. This is true for all coarse-graining volumes larger than an unit cell. What is
the effect of these configurations on elasticity and how do they influence mechanical behavior? Recently, there has been significant progress in the
study of non-affineness in solid plasticity - especially in the context of rheological properties of amorphous materials and granular solids which
show jamming behavior \cite{FALK_LA,LEMAI,MALON}. Localized non-affine regions consisting of particles capable of large reorientations have been shown to be involved 
in relaxation processes in these systems. Is there an analog of such regions in an ideal, crystalline solid?  
A study of these fluctuations in ideal solids, as presented in section \ref{NAFSTAT},  may help us understand complex dynamics in solids better. 

The organization of this paper, together with a short summary of our main results is as follows. In section \ref{LTHEO} we derive an 
analytic form of the non-local elastic response function, or compliance $\chi_{ij}(\vec{r},\vec{r}^{\prime})$ ($i = x,y$),  which is defined 
as the strain $\varepsilon_{ij}(\vec{r}^{\prime})$ produced at position $\vec{r}^{\prime}$ due to a stress 
$\sigma_{ij}(\vec{r})$ at $\vec{r}$. In order to do this, we consider a Landau expansion \cite{CHAIKI} of the free energy in terms of the strains, 
keeping up to quartic terms in the gradients. Next in section \ref{SIMUS} we present Monte Carlo computer simulations of a harmonic crystal. The 
calculation of the local strain field corresponds to a coarse-graining procedure and allows us to construct the strain-strain correlation 
function, $G_{ij}$ which is related to the response function via  $\chi_{ij} = (k_BT)^{-1} G_{ij}$. For a homogeneous solid without external load (i.e.
$\langle\varepsilon_{i}(\vec{r})\rangle = 0$) the correlation functions 
are given by $G_{ij} (\vec{r}^{\prime}) = V<\varepsilon_{i}(\vec{0})\,\varepsilon_{j}(\vec{r}^{\prime})>$. The $<...>$ denote a thermal average 
(and in addition one over the choice of origin) and $k_B T$ is the Boltzmann constant times the temperature. We compare our results, obtained 
for a variety of ensembles and boundary conditions to that of the Landau theory. A common feature in experimental systems is the presence of an embedding medium, surrounding the analyzed region of the sample. The effects of such an embedding medium on
the strain correlations are discussed and visualized by use of a statistical sum rule. 
One of our significant results is that though the forms of the correlation functions and 
their limiting values as predicted by the Landau theory are reproduced, the $G_{ij}$ obtained from simulations through our coarse-graining procedure 
differ by an additional background contribution which is not negligible. In section \ref{NAFSTAT} we argue that this is a consequence of 
non-affine displacements, which are not considered in the ansatz for the Landau theory. The amount of non-affinity in a given configuration can be quantified by calculating $\chi$, the deviation of the actual configuration from one obtained from an affine
 transformation of the reference lattice. We analyze this non-affineness in detail and show that the probability distribution of 
 the non-affine field $P(\chi)$ can  be computed within a simple ``cell model'' approximation. $P(\chi)$ shows well defined scaling properties 
 with the spring stiffness $f$ and the coarse-graining length $\Lambda$. The auto-correlation function for $\chi$ is shown to be short ranged 
 decaying rapidly for distances much larger than the coarse-graining length. Finally, we conclude our paper indicating future directions for research. 

\section{Landau theory for the strains\label{LTHEO}}
The two dimensional elastic continuum described by the linear elastic Hamiltonian, 
\begin{equation}
\beta {\cal H}_{0} = \frac{1}{2} \int d{\vec r}\,\, C_{ijkl}\epsilon_{ij}\epsilon_{kl}
\label{h2d}
\end{equation}
is perpetually in a critical state \cite{CHAIKI}. The displacement correlations $\langle {\vec u}({\vec r}) \cdot {\vec u}({\vec r}\,') \rangle$ decay algebraically and the solid shows 
quasi-long ranged order with the elastic susceptibilities diverging logarithmically with system size $L$. For all practical purposes, however, this weak 
divergence may be ignored and non-zero elastic moduli may be defined and computed. In real solids, an upper length scale cutoff is set by the 
typical distance between defect pairs. 

The fact that the solid state is critical also implies that the Hamiltonian ${\cal H}_{0}$ in Eq.(\ref{h2d}) is a fixed point Hamiltonian which should be 
invariant under a coarse- graining procedure, 
unless topological defects such as dislocations are present. This, again, is not strictly true, as we shall 
demonstrate in section \ref{SIMUS}. This is because in a molecular system, any Hamiltonian such as ${\cal H}_0$ is realized only in an approximate, 
discretized sense, the displacements ${\vec u}$ being carried by the individual particles. Anticipating some of our results in section \ref{SIMUS}, we 
use the following dimensionless Landau functional \cite{CHAIKI,ERINGE}:
\begin{eqnarray}
\beta\mathcal{F} &=& \frac{1}{2}\int d^{2}x~\sum_{i=1}^{3}\left\{a_{i}e_{i}^{2}+c_{i}(\nabla e_{i})^{2}+c'_{i}(\nabla^{2} e_{i})^{2}\right\}\label{LANDU}
\end{eqnarray}
Here with ($i=1-3$) the dimensionless constants $a_{i}$ are the elastic moduli of the system, while $c_{i}$ and $c'_{i}$ are phenomenological 
coefficients and the three strains $e_i$ are given by:
$e_{1}=\epsilon_{xx}+\epsilon_{yy}$, describing pure volume changes; $e_{2}=\epsilon_{xx} -\epsilon_{yy}$, 
describing deviatoric shear strains and  $e_{3}=\frac{1}{2}(\epsilon_{xy}+\epsilon_{yx})$,  describing pure shear strains. 
The phenomenological coefficients $c_{i}$ have the dimension of a $\textrm{length}^{2}$. Thus we interpret $\xi_{el,i} \sim
 \sqrt{c_{i}}$ as a correlation length. Note that unless noted otherwise throughout the paper all 
lengths are given in units of $a$, the lattice parameter of the underlying triangular lattice in the simulated systems.

The terms quadratic in the strains $e_{i}$ represent the local part in this ansatz. Non-local contributions are included via the gradient terms. Note that 
the Landau functional Eq.(\ref{LANDU}) should be strictly valid for excitations of wavelengths longer than a short-wavelength cutoff. Short wavelength 
excitations are suppressed by the gradient terms \cite{CHAIKI,ERINGE} and we have ignored the possibility of defects. 

For non-uniform strains, spatial fluctuations of the strains couple and only one of the strain variables $e_{i}$ in this ansatz is an independent variable as we show below. Firstly, 
all forces within and on the system
must cancel for the system to be in thermodynamic equilibrium, i.e. $\frac{\partial \sigma_{ij}}{\partial x_{j}}=0$ for a solid under zero external stress; the case of external hydrostatic stress is presented in section \ref{PRESS}. The stress
tensor $\sigma_{ij}$ can be obtained directly from Eq.(\ref{LANDU}), as $\sigma_{ij}=\frac{\delta~\beta\mathcal{F}}{\delta \epsilon_{ij}}$ . 
For a two-dimensional crystal this condition reads in Fourier space:
\begin{eqnarray}
&&k_{x}a_{1}\tilde{e}_{1}+k_{x}a_{2}\tilde{e}_{2}+k_{y}a_{3}\tilde{e}_{3} = 0\label{GLGW1} \\
&&k_{y}a_{1}\tilde{e}_{1}-k_{y}a_{2}\tilde{e}_{2}+k_{x}a_{3}\tilde{e}_{3} = 0\label{GLGW2} 
\end{eqnarray}

In addition St Venant's compatibility condition must be considered in the calculations, which ensures an unique relation between the displacement field
$\vec{u}$ and the strain fields $\epsilon_{ij}$. For a two-dimensional crystal this
simplifies to 
\begin{equation}
-k^{2}\tilde{e}_{1}+(k_{x}^{2}-k_{y}^{2})\tilde{e}_{2}+4k_{x}k_{y}\tilde{e}_{3}=0\label{STV}
\end{equation}
in Fourier space. With the help of these three conditions the kernels $\tilde{Q}_{ij}$ relating the strain variables
$\tilde{e}_{i}=\tilde{Q}_{ij}\tilde{e}_{j}$ can be derived. The resulting relations are given in detail below:
\begin{eqnarray}\nonumber
\tilde{e}_{2}&=&-\left(\frac{4a_{1}+2a_{3}}{a_{1}+a_{2}}\right)\left(\frac{k_{x}k_{y}}{k_{x}^{2}-k_{y}^{2}}\right)\tilde{e}_{3}=\tilde{Q}_{23}\tilde{e}_{3}\\\nonumber
\tilde{e}_{1}&=&-\left(\frac{2a_{3}-4a_{2}}{a_{1}+a_{2}}\right)\left(\frac{k_{x}k_{y}}{k^{2}}\right)\tilde{e}_{3}=\tilde{Q}_{13}\tilde{e}_{3}\\\nonumber
\tilde{e}_{1}&=&\left(\frac{a_{3}-2a_{2}}{2a_{1}+a_{3}}\right)\left(\frac{k_{x}^{2}-k_{y}^{2}}{k^{2}}\right)\tilde{e}_{2}=\tilde{Q}_{12}\tilde{e}_{2}\nonumber
\end{eqnarray}
Note that the properties of the correlation functions are set by the wave vector dependence of the kernels $\tilde{Q}_{ij}$.

We shall also need kernels relating the strains $e_{i}$ to the local, microscopic
rotations given by the anti-symmetric part of the strain tensor 
$\theta=\left(\partial u_{y}/\partial x - \partial u_{x}/\partial y\right)/2$. Local rotations
$\theta$ are related to the deviatoric strains $e_{2}$ and the pure shear strains $e_{3}$. These strains are coupled in case of the presence of a surrounding, 
embedding medium. Starting from the definitions of the strain variables the partial derivative $\partial^{3}\theta/\partial x^{2}\partial y$ 
can be expressed solely in terms of $e_{2}$, $e_{3}$ and $\theta$. Thus one obtains 
the following relation between the three shear strain variables in Fourier space: 
\begin{eqnarray}\nonumber
\tilde{\theta}(k_{x}^{2}+k_{y}^{2})=(k_{x}^{2}-k_{y}^{2})\tilde{e}_{3}-k_{x}k_{y}\tilde{e}_{2}\label{theta_1}\nonumber
\end{eqnarray}
We may now derive a new kernel relating  $\tilde{\theta}$ to e.g. the pure shear strain $\tilde{e}_{3}$ using the previously derived kernel $\tilde{Q}_{23}$. 
This results in:
\begin{eqnarray}\nonumber
\tilde{e}_{3}=\frac{k_{x}^{4}-k_{y}^{4}}{(k_{x}^{2}-k_{y}^{2})^{2}+k_{x}^{2}k_{y}^{2}\left(\frac{4a_{1}+2a_{3}}{a_{1}+a_{2}}\right)}~ \tilde{\theta} = \tilde{Q}_{3\theta}~ \tilde{\theta}
\end{eqnarray}
As the considerations in section \ref{EMBED} will show, it will be helpful to define the strain variable $e_{2\theta}=2\theta$ and analyze its two-point correlation
function in embedded systems, such as the colloidal crystal discussed in \cite{KFRANZ2}. 

Switching to a discretized notation, appropriate for comparison with our simulations,  so that $e_{1}(\vec{x})\rightarrow e_{1,\vec{m}}$ and $\vec{x}\rightarrow \vec{R}_{\vec{m}}$, where
$\vec{m}$ (a tuple of integer lattice indices) specifies the position on a coarse-graining, square mesh the free energy can be rewritten as,
\begin{eqnarray}\nonumber
\beta \mathcal{F} &=&\frac{v_{z}}{2}\sum_{\vec{m}}\sum_{i=1}^{3}\left(a_{i}e_{i,\vec{m}}^{2}+c_{i}(\nabla e_{i,\vec{m}})^{2}+c'_{i}(\nabla^{2}e_{i,\vec{m}})^{2}\right)\nonumber
\end{eqnarray}
Here $v_{z}=V_{z}/a^{2}$ is the dimensionless volume of the coarse-graining cell and the summation runs over all $N$ cells. 
One may use the above relations to finally express $\beta\mathcal{F}$ as a harmonic functional of only one of the strain components. This allows 
the direct calculation of the analytic form of the two-point correlation functions, as the
partition function factorizes (e.g. \cite{GOLDF}). Here we choose as an example the strains $\tilde{e}_{1}$ and obtain the following expression 
for the functional, written as a sum over the wave vectors:
\begin{widetext}
\begin{eqnarray}\nonumber
\beta\mathcal{F}&=&\frac{1}{2v}\sum_{\vec{k}}\{a_{1}+k^{2}c_{1}+k^{4}c'_{1}+(a_{2}+c_{2}k^{2}+c'_{2}k^{4})(\tilde{Q}_{21}(\vec{k}))^{2}+(a_{3}+c_{3}k^{2}+c'_{3}k^{4})(\tilde{Q}_{31}(\vec{k}))^{2}\}\tilde{e}_{1,\vec{k}}\tilde{e}_{1,\vec{k}}^{*}
\end{eqnarray}
\end{widetext}
with $v=V/a^{2}$, the dimensionless volume of the system. It is now straight forward to calculate the two-point correlation function starting from its general definition and taking into account the fact, that 
the average strains $\langle e_{i}(\vec{r})\rangle$ in a crystal, which is under no load, is zero.
\begin{eqnarray}\nonumber
G_{ii}(\vec{r},\vec{r}^{\prime})&=&\langle e_{i}(\vec{r})e_{i}(\vec{r}^{\prime})\rangle-\langle e_{i}(\vec{r})\rangle\langle
e_{i}(\vec{r}^{\prime})\rangle\\\nonumber
&=&\frac{1}{v^{2}}\sum_{\vec{k}\vec{k'}}e^{i((\vec{k}\cdot\vec{r})(\vec{k}^{\prime}\cdot\vec{r}^{\prime}))}\langle
\tilde{e}_{\vec{k}}\tilde{e}_{\vec{k}^{\prime}}\rangle\\\nonumber
\end{eqnarray}
The calculation yields the following relation for the average of the Fourier coefficients:
\begin{eqnarray}\nonumber
\langle \tilde{e}_{\vec{k}}\tilde{e}_{\vec{k}'}\rangle = \delta_{\vec{k}+\vec{k}^{\prime},\vec{0}}\langle |\tilde{e}_{\vec{k}}|^{2}\rangle=
\delta_{\vec{k}+\vec{k}',\vec{0}}v/A(\vec{k})
\end{eqnarray}
Where,
$A(\vec{k})=a_{1}+k^{2}c_{1}+k^{4}c'_{1}+(a_{2}+c_{2}k^{2}+c'_{2}k^{4})(\tilde{Q}_{21}(\vec{k}))^{2}+(a_{3}+c_{3}k^{2}+c'_{3}k^{4})(\tilde{Q}_{31}(\vec{k}))^{2}$.
 As $\tilde{G}_{ii}(\vec{k})= \langle |\tilde{e}_{\vec{k}}|^{2}\rangle/v$ one can now write down the analytic form of the strain-strain correlation
functions or rather their inverse in detail. For $i=1,2,3$ the structure of the strain-strain correlation functions is the same, while for $i=2\theta$ it differs slightly:
\begin{subequations}
\begin{eqnarray}
\tilde{G}_{ii}(\vec{k}\neq0)^{-1} &=&
a_{i}+k^{2}c_{i}+k^{4}c'_{i}\\\nonumber 
&&+\sum_{j\neq i,j=1}^{3}(a_{j}+c_{j}k^{2}+c'_{j}k^{4})(\tilde{Q}_{j1}(\vec{k}))^{2}\\\nonumber
\tilde{G}_{ii}(\vec{0})^{-1} &=& a_{i}\\\nonumber
\end{eqnarray}
\begin{eqnarray}
\tilde{G}_{2\theta2\theta}(\vec{k}\neq0)^{-1}&=&\Big(a_{3}+c_{3}k^{2}+c_{3}^{'}k^{4}\\\nonumber
&+&\sum_{j=1}^{2}(a_{j}+c_{j}k^{2}+c_{j}^{'}k^{4})(\tilde{Q}_{j3})^{2}\Big)(\tilde{Q}_{3~2\theta})^{2}\\
\tilde{G}_{2\theta2\theta}(\vec{0})^{-1}&=&\frac{a_{3}}{4}\nonumber
\end{eqnarray}
\label{ANACOR}
\end{subequations}

\subsection{Properties of the strain-strain correlation functions}
\begin{figure}
\begin{center}
\includegraphics[width=7.25cm]{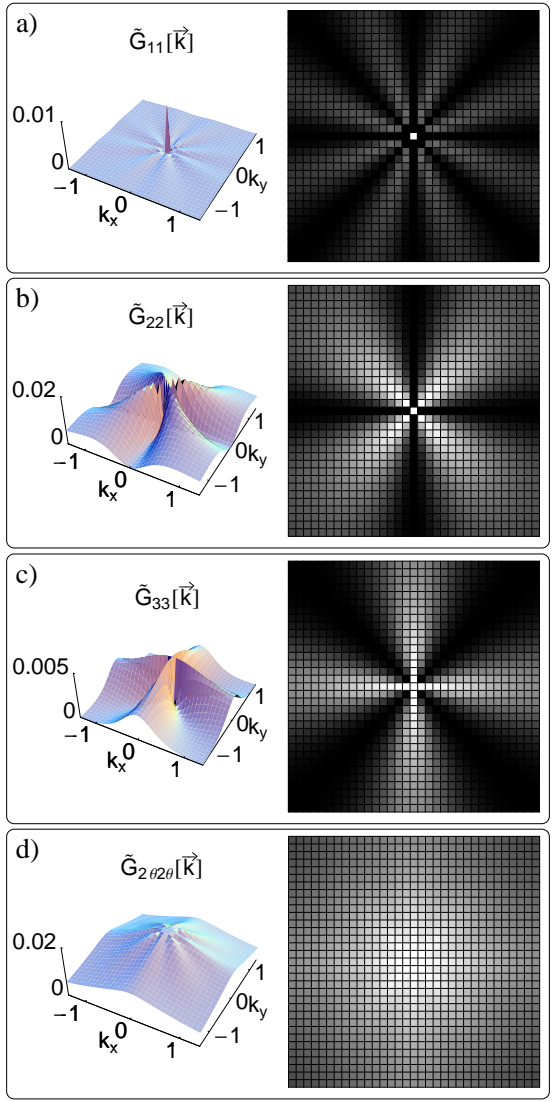}
\caption{(Color online) The analytic form of the strain-strain correlation functions. For each function a surface plot and next to it a density plot
are shown. In the density plot the maxima are white, minima are black. The set of parameters used for plotting was obtained from a
Monte Carlo simulation of a harmonic triangular lattice in the $NVT$ ensemble with periodic boundary conditions: $a_{1}=97.4$, $a_{2}=48.1$, $a_{3}=198.2$, $c_{1}=54.3$,
 $c_{2}=37.4$, $c_{3}=118.2$, $c_{1}'=-86.1$, $c_{2}'=-1.3$ and $c_{3}'=-18.8$. a) $G_{11}(\vec{k})$, b) $G_{22}(\vec{k})$ , c)
$G_{33}(\vec{k})$ and d) $G_{2\theta 2\theta}(\vec{k})$.\label{a_plots_kont}}
\end{center}
\end{figure}

The analytic strain-strain correlation functions $\tilde{G}_{ii}(\vec{k})$ (the inverse of which were given in Eq.(\ref{ANACOR})) are plotted in figure \ref{a_plots_kont}. The set of parameters $a_{i}$, $c_{i}$ and $c'_{i}$ used in figure \ref{a_plots_kont} were obtained from Monte-Carlo simulations in the $NVT$ ensemble with
periodic boundary conditions of a harmonic triangular lattice, to be discussed in section \ref{NVTPBC}.  While the deviatoric and shear strain correlation functions $\tilde{G}_{22}$ and $\tilde{G}_{33}$ have four-fold
symmetries, the correlation function of the dilatation $\tilde{G}_{11}$ has an eight-fold symmetry. The correlation functions may be interpreted as the response 
of the system to a localized perturbation at the origin. This perturbation is either a dilatation, a deviatoric shear or a pure shear. The deformation 
of the solid may be decomposed as a superposition of the eigenmodes of the system. The eigenmodes for a square box, are plane waves with polarizations either longitudinal or transverse to the coordinate axes with the eigenfrequencies forming a discrete spectrum: 
$\omega_{nm}=\frac{2\pi}{L}c_{\alpha}\sqrt{n^{2}+m^{2}}$ with $n,m\in N_{0}$. 
Thus the wave vector of the eigenmodes along the diagonal, i.e. ($n=m$), exhibits a four-fold degeneracy, while those parallel to the coordinate axis, i.e. $n\neq m$ and $n\neq m \neq 0$,  have an eight-fold degeneracy. A local dilatation as perturbation results in a superposition of eigenmodes with four- as well as  eight-fold
degeneracy. This leads to the eight-fold rotational symmetry visible for the strain correlation function $\tilde{G}_{11}$. In contrast to this the two possible shear perturbations will excite elastic waves that are superpositions of \emph{exclusively} eigenmodes with four-fold degeneracy. For this reason the corresponding correlation functions $\tilde{G}_{22}$ and $\tilde{G}_{33}$ exhibit only a four-fold rotational
symmetry.
As was discussed in \cite{KFRANZ3}, the presence of defects, breaks the rotational symmetries of the strain correlation functions.

The details of the structure of the correlation functions are dominated by the dependence of the kernels $\tilde{Q}_{ij}$ on 
the wave vector $\vec{k}$, especially the cases $k_{x}=k_{y}$, $k_{x}\rightarrow 0$ while $k_{y}\neq 0$  and $k_{y}\rightarrow 0$ 
while $k_{x}\neq 0$. In particular, we obtain the following relations:
\begin{eqnarray}\nonumber
\tilde{Q}_{12} &=& \left\{\begin{array}{cl}
0&~~~~~\textrm{, for}~k_{x}=k_{y}\\
-\left(\frac{a_{3}-2a_{2}}{2a_{1}+a_{3}}\right)&~~
~~~\textrm{, for}~k_{x}\rightarrow 0~~\textrm{,}~~k_{y}\neq 0\\
\left(\frac{a_{3}-2a_{2}}{2a_{1}+a_{3}}\right)&~~~~~\textrm{, for}~k_{y}\rightarrow
0~~\textrm{,}~~k_{x}\neq 0\\
\end{array}\right. \\\nonumber
\tilde{Q}_{13} &=& \left\{\begin{array}{cl}
\left(\frac{2a_{2}-a_{3}}{a_{1}+a_{2}}\right)&~~\textrm{, for}~k_{x}=k_{y}\\
0&~~\textrm{, for}~k_{x}\rightarrow 0~~\textrm{,}~~k_{y}\neq 0\\
0&~~\textrm{, for}~k_{y}\rightarrow 0~~\textrm{,}~~k_{x}\neq 0\\
\end{array}\right.\\\nonumber
\tilde{Q}_{23} &=& \left\{\begin{array}{cl}
\infty&~~\textrm{, for}~k_{x}=k_{y}\\
0&~~\textrm{, for}~k_{x}\rightarrow 0~~\textrm{,}~~k_{y}\neq 0\\
0&~~\textrm{, for}~k_{y}\rightarrow 0~~\textrm{,}~~k_{x}\neq 0\\
\end{array}\right.
\end{eqnarray}
For the kernel $\tilde{Q}_{3~2\theta}$ relating $e_{3}$ to $e_{2\theta}$ the behavior along the specific directions in Fourier
space for $\vec{k}\rightarrow \vec{0}$ can be extracted from the behavior of the correlation function $\tilde{G}_{33}(\vec{k})$ 
and $\tilde{G}_{22}(\vec{k})$.
Along the coordinate axis the kernel relating $\tilde{e}_{3}$ to $\tilde{e}_{2\theta}$ becomes a constant.
\begin{eqnarray}\nonumber
\tilde{Q}_{3~2\theta} &=& \left\{\begin{array}{cl}
-1/2&\textrm{for}~k_{x}\rightarrow 0~~\textrm{,}~~k_{y}\neq 0\\
~~1/2&\textrm{for}~k_{y}\rightarrow 0~~\textrm{,}~~k_{x}\neq 0\\
\end{array}\right.
\end{eqnarray}
Thus the continuous behavior of  the correlation function $\tilde{G}_{33}(\vec{k})$ for $\vec{k}\rightarrow 0$
\emph{along} the coordinate axis carries over to the correlation function $\tilde{G}_{2\theta 2\theta}(\vec{k})$.
The behavior along the diagonals $k_{x}=k_{y}$ can be extracted from the behavior of the product of the kernels $\tilde{Q}_{23}^{2}\tilde{Q}_{3~2\theta}^{2}$, which can be
shown to equal $1$. 
Upon insertion of these relations into the equations for the inverse of the correlation functions their behavior for these limiting cases can be extracted:
\begin{eqnarray}\nonumber
\tilde{G}^{-1}_{11}(\vec{k})&=&
\left\{\begin{array}{cl}
~~~~~~~~~\infty&~~~~~~~~~\textrm{,}~~k_{x}=k_{y}\neq 0\\
~~~~~~~~~\infty&~~~~~~~~~\textrm{,}~~k_{x}\rightarrow 0~~\textrm{,}~~k_{y}\neq 0\\
~~~~~~~~~\infty&~~~~~~~~~\textrm{,}~~k_{x}\rightarrow 0~~\textrm{,}~~k_{y}\neq 0\\
\end{array}\right.
\end{eqnarray}
\begin{eqnarray}\nonumber
\tilde{G}^{-1}_{22}(\vec{k})&=&
\left\{\begin{array}{cl}
a_{2}+c_{2}k^{2}+c_{2}'k^{4}&\textrm{,}~~k_{x}=k_{y}\neq 0\\
\infty&\textrm{,}~~k_{x}\rightarrow 0~~\textrm{,}~~k_{y}\neq 0\\
\infty&\textrm{,}~~k_{x}\rightarrow 0~~\textrm{,}~~k_{y}\neq 0\\
\end{array}\right.
\end{eqnarray}
\begin{eqnarray}\nonumber
\tilde{G}^{-1}_{33}(\vec{k})&=&
\left\{\begin{array}{cl}
\infty&\textrm{,}~~k_{x}=k_{y}\neq 0\\
a_{3}+c_{3}k^{2}+c_{3}'k^{4}&\textrm{,}~~k_{x}\rightarrow 0~~\textrm{,}~~k_{y}\neq 0\\
a_{3}+c_{3}k^{2}+c_{3}'k^{4}&\textrm{,}~~k_{y}\rightarrow 0~~\textrm{,}~~k_{x}\neq 0\\
\end{array}\right.
\end{eqnarray}
\begin{eqnarray}\nonumber
\tilde{G}^{-1}_{2\theta 2\theta}(\vec{k})&=&
\left\{\begin{array}{cl}
a_{2}+c_{2}k^{2}+c_{2}'k^{4}&\textrm{,}~~k_{x}=k_{y}\neq 0\\
(a_{3}+c_{3}k^{2}+c_{3}'k^{4})/4&\textrm{,}~~k_{x}\rightarrow 0~~\textrm{,}~~k_{y}\neq 0\\
(a_{3}+c_{3}k^{2}+c_{3}'k^{4})/4&\textrm{,}~~k_{x}\rightarrow 0~~\textrm{,}~~k_{y}\neq 0\\
\end{array}\right.
\end{eqnarray}
These considerations show, that in certain directions in Fourier space the shear strain variables become independent from each other 
and are continuous for $\vec{k}\rightarrow\vec{0}$ . These are the directions, along which a fit
will give direct access to the elastic constants and correlation lengths of the system.
\begin{figure}[t!]
\begin{center}
\includegraphics[angle=-0,width=6.75cm]{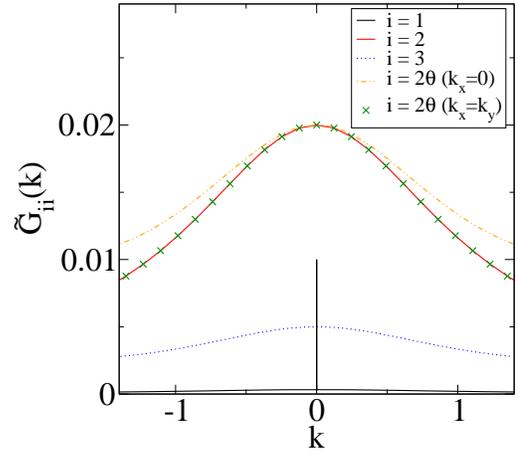}
\caption{(Color online) Cuts of the analytic strain-strain correlation functions along specific directions in Fourier space. The set of parameters used for plotting 
is: $a_{1}=100$, $a_{2}=50$, $a_{3}=200$ and $c_{1}=54.3$, $c_{2}=37.4$, $c_{3}=118.2$, $c_{1}'=-86.1$, $c_{2}'=-1.3$, $c_{3}'=-18.8$.
The correlations lengths were obtained from a Monte Carlo simulation of a harmonic triangular lattice in the $NVT$ ensemble with 
periodic boundary conditions. $\tilde{G}_{11}(\vec{k})$ is shown along the direction $k_{y}=2k_{x}$, 
$\tilde{G}_{22}(\vec{k})$ along the diagonal $k_{x}=k_{y}$, $\tilde{G}_{33}(\vec{k})$ along the $k_{y}$-axis and $\tilde{G}_{2\theta
2\theta}(\vec{k})$ along the diagonal $k_{x}=k_{y}$ and along the $k_{y}$-axis.\label{a_cuts2}}
\end{center}
\end{figure} 
Figure \ref{a_cuts2} shows cuts in Fourier space which correspond to these specific directions for the strain correlation functions
$\tilde{G}_{ii}(\vec{k})$. For the correlation function of the pure shear strain $\tilde{G}_{22}(\vec{k})$ a cut along the diagonals is
shown, while for the deviatoric strain correlation function $\tilde{G}_{33}(\vec{k})$ a cut along the coordinate axis is shown. The strain
correlation function  for the dilatation $\tilde{G}_{11}(\vec{k})$ in contrast is not continuous for $\vec{k}\rightarrow\vec{0}$. If one
considers for example the direction $k_{y}=2k_{x}$ the kernels $\tilde{Q}_{ij}$ relating the strain variables turn into constant weighting
factors: $\tilde{Q}_{21}=-\left(\frac{2a_{1}+a_{3}}{a_{3}-2a_{2}}\right)\left(\frac{5}{3}\right)$ and 
$\tilde{Q}_{31}=\left(\frac{a_{1}+a_{2}}{4a_{2}-2a_{3}}\right)\left(\frac{5}{2}\right)$. 
Thus along the direction $k_{y}=2k_{x}$ the inverse of the strain correlation function for the dilatation $\tilde{G}_{11}^{-1}(\vec{k})$ for $k\rightarrow 0$ is given by
\begin{eqnarray}\nonumber
\tilde{G}^{-1}_{11}(\vec{k}\rightarrow 0)&=&
\Bigg(a_{1}+a_{2}\left[-\left(\frac{2a_{1}+a_{3}}{a_{3}-2a_{2}}\right)\left(\frac{5}{3}\right)\right]^{2}\\\nonumber
&&~+ a_{3}\left[\left(\frac{a_{1}+a_{2}}{4a_{2}-2a_{3}}\right)\left(\frac{5}{2}\right)\right]^{2}\Bigg)\\\nonumber
&\neq& \\\nonumber
\tilde{G}^{-1}_{11}(\vec{k}=0)&=&a_{1}
\end{eqnarray} 
So the strain correlation function for the dilatation $\tilde{G}_{11}(\vec{k})$ exhibits a pronounced discontinuity for $\vec{k}\rightarrow \vec{0}$. In order to
illustrate this fact, consider the set of parameters used in the simulations of a harmonic triangular lattice (to be discussed in section \ref{SIMUS}). The choice of the spring constant sets the elastic constants of the system under consideration to $a_{1}=100$, $a_{2}=50$ and $a_{3}=200$. For $\vec{k}\rightarrow 0$ one has $\tilde{G}^{-1}_{11}(\vec{k}\rightarrow 0)\approx 3025$, which is approximately $30$ times as much as the value for  $\vec{k}=0$, i.e. $\tilde{G}^{-1}_{11}(\vec{k}=0)=a_{1} = 100$, set by the bulk modulus --- showing that non-uniform dilations tend to be severely penalized in this solid. 

Nevertheless, provided that the value of the bulk modulus is determined e.g. from
$\tilde{G}_{11}(\vec{k}=\vec{0})$  the coefficients $c_{1}$ and $c_{1}'$ can also be obtained by fitting one of the correlation functions along a cut in
Fourier space. So in principle all $9$ parameters of the free energy functional can be determined from an analysis of the strain-strain correlation functions. 
Like the correlation function of the dilatation $\tilde{G}_{11}(\vec{k})$ the correlation function of the microscopic
rotations shows an eight-fold rotational symmetry. Unlike $\tilde{G}_{11}(\vec{k})$, however,  $\tilde{G}_{2\theta 2\theta}(\vec{k})$ is continuous for $\vec{k}\rightarrow \vec{0}$
along the coordinate axis and the diagonal (compare figure \ref{a_cuts2}). Therefore fits along these directions can be used to determine the elastic
constants $a_{2}$ and $a_{3}$ as well as the coefficients $c_{2}$, $c_{2}'$ and $c_{3}$, $c_{3}'$. 

We shall next discuss the result of a coarse-graining procedure, which attempts to obtain these correlation functions and therefore the parameters of 
the Landau free energy functional from  Monte Carlo simulations of the harmonic lattice. The coefficients of all the second and fourth order terms 
involving gradients of strain are found to be non-vanishing showing that coarse-graining generates these  
higher order non-local terms in the free energy.

\section{Monte Carlo simulations of a harmonic crystal\label{SIMUS}}
The analysis of a harmonic crystal is convenient for a comparison with the results of the Landau theory presented in the last section, since  
the elastic moduli can be directly calculated from the spring constants. We consider a harmonic triangular lattice with a 
Hamiltonian $\mathcal{H}=k_{B}T(f/2)\sum_{m,n=1}^{N}(|\vec{r}_{m}-\vec{r}_{n}|-a)^{2}$ where $f$ is
the spring constant and $a$ the lattice parameter of the triangular lattice. The elastic moduli are related to the spring constant $f$
via: $K=a_{1}=(\sqrt{3}/2)f$, $\mu=a_{2}=(\sqrt{3}/4)f$ and $4\mu=a_{3}=\sqrt{3}f$. 
Furthermore the harmonic triangular lattice has been shown to be a successful model for the interpretation of
experiments on colloidal crystals \cite{KEIM_HARMONIC}. 
It is modeled by $N$ point-particles each of them hard-wired by spring constants $f$ to the six nearest
neighbors. We have carried out Monte Carlo simulations in the constant $NpT$ and $NVT$ ensembles with periodic boundary conditions. We also 
mention briefly results for a system with open boundary conditions which were presented elsewhere \cite{KFRANZ2}. Next the influence of hydrostatic pressure is
analyzed by Monte Carlo simulations in the constant $NpT$ ensembles with periodic boundary conditions. Finally we consider the 
effect of a surrounding elastic medium and finite size effects in order to make contact with experiments on colloids.  

The knowledge of the configurations and the reference lattice allows for a direct calculation of the displacement field $\vec{u}(\vec{r})$. In order to calculate the
corresponding strain field partial differentials of the displacement field have to be calculated. We follow the procedure by Falk and Langer
\cite{FALK_LA} and 
calculate the strain field by minimizing the error in the affine transformation that relates the actual configuration $\{\vec{r}\}$ to the reference lattice $\{\vec{R}\}$.
\begin{eqnarray}\nonumber
\vec{r}=\vec{R}+\vec{u}(\vec{R})=(\mathbf{1}+\mathbf{\epsilon})\vec{R}
\end{eqnarray}
The mean-squared error in this mapping $\chi$ is thus a measure of how well the given situation can be described within the framework of linear elasticity theory
and quantifies the non-affinity of the given displacement field. Falk and Langer \cite{FALK_LA} analyzed the temporal development of strains. Here we use an analogous
definition for $\chi$ in thermodynamics equilibrium, evaluating the strains and non-affineness with respect to the reference lattice:
\begin{eqnarray}\nonumber
\chi(\vec{r}_{0})=\sum_{m=1}^{N_{B}}\sum_{i=1}^{2}\left(r_{m}^{i}-r_{0}^{i}-\sum_{j=1}^{2}(\delta_{ij}+\epsilon_{ij})[R_{m}^{j}-R_{0}^{j}]\right)^{2}
\label{CHIDEF}
\end{eqnarray}
Here $\vec{r}_{0}$ is the position, at which the strains are to be calculated, and $N_{B}$ is the number of neighboring particles considered. 
This corresponds to a coarse-graining procedure, in which $N_{B}$ is set by the choice of coarse-graining length $\Lambda$, i.e. cutoff radius within
which particles are considered in the calculation. For the results presented in this section, we have used a cutoff radius of $\Lambda=1.3$ resulting in
$N_{B} = 6$. In section \ref{NAFSTAT} we present some systematics showing how some of our results depend on the coarse-graining length $\Lambda$. 

In the calculation of the strain-strain correlation functions a second coarse-graining step is employed, when mapping the triangular lattice to a square mesh.
This facilitates the numerical Fourier transformation of the calculated real space correlation functions. The wave vectors are limited to the first Brillouin zone, i.e. $k_{j}\in[-\frac{\pi}{l_{m}},\frac{\pi}{l_{m}}]$, with $j=x,y$. Here $l_{m}$ represents the lattice parameter of 
the coarse-grained, square mesh. 
Care must be taken in the choice of $l_{m}$ to keep the coarse-graining volume large enough so 
that artifacts due to the discreteness of the triangular lattice (and insufficient averaging) are avoided. In most of the results presented here we 
used $l_{m}=2.25a$, where $a$ is the lattice parameter of the original, triangular lattice. Lastly, one also needs to be careful about 
correcting for global rotations and translations of the lattice especially for the case of open boundary conditions so as not to introduce artificial sources of error.   

Simulations of the harmonic triangular lattice were done for three system sizes $N=3120$, $4736$ and $5822$. We first discuss the results for simulations with spring
constant $\beta a^{2}f=200/\sqrt{3}$ in the $NpT$ ensemble with periodic boundary conditions (and external pressure $p = 0$) and the $NVT$ ensemble with periodic boundary 
conditions. A crystal with open boundary conditions was discussed in detail in \cite{KFRANZ2} and will only be mentioned briefly. In what follows we analyze the 
influence of a hydrostatic, external pressure and of a surrounding, embedding medium.

\subsection{$NpT$ ensemble with periodic boundary conditions\label{NPTPBC}}
\begin{table}
\begin{center}
\begin{tabular}{|l|r|r|r|}
\hline
~& $a_{1}$& $a_{2}$& $a_{3}$\\
\hline
~&~&~&~\\
calculated from $f$ &$100~/\beta a^{2}$&$50~/\beta a^{2}$&$200~/\beta a^{2}$\\
\hline
from fluctuations&~&~&~\\
of $\mathbf{h}$&$98.9~/\beta a^{2}$&$49.4~/\beta a^{2}$&$196.7~/\beta a^{2}$\\
\hline
~&~&~&~\\
from $\tilde{G}(\vec{k}=\vec{0})$&$96.8~/\beta a^{2}$&$48.6~/\beta a^{2}$&$190.8~/\beta a^{2}$\\
\hline
~&~&~&~\\
from fits of $\tilde{G}(\vec{k}\neq \vec{0})$&-&$49.1~/\beta a^{2}$&$195.7~/\beta a^{2}$\\
\hline
\end{tabular}
\end{center}
\caption{A comparison of the elastic constants as calculated from the spring constant $f$ and as obtained by use of various methods 
from data of a Monte Carlo simulation of a harmonic triangular lattice with $N=3120$ particles in the $NpT$ ensemble at zero 
hydrostatic pressure $\beta a^{2} p =0.0$ and spring constant $\beta a^{2} f= 200/\sqrt{3}$. \label{PRR_tabelle}}
\end{table}
\begin{figure}
\begin{center}
\includegraphics[width=7.25cm]{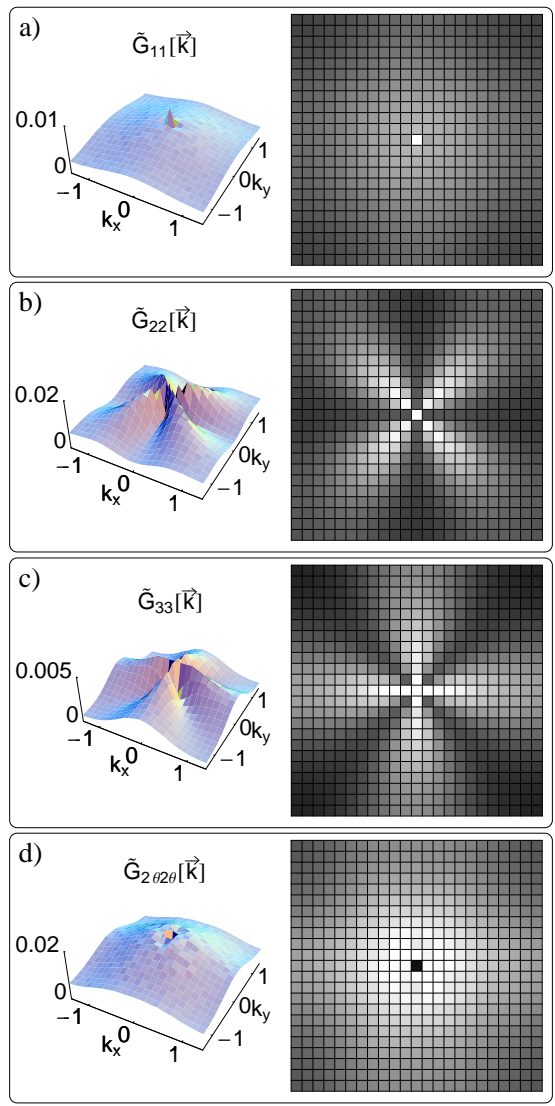}
\caption{(Color online) Strain-strain correlation functions of a harmonic triangular lattice at zero hydrostatic pressure as obtained from Monte
Carlo simulations in the $NpT$ ensemble with periodic boundary conditions. Results for a system with  $N=3120$ particles and a
spring constant $\beta a^{2} f= 200/\sqrt{3}$ are shown. For each function a surface plot and next to it a density plot
are displayed. In the density plot the maxima are white, minima are black.\label{npt_ort} }
\end{center}
\end{figure}
For the simulations of the harmonic crystal in the $NpT$ ensemble we use the algorithm of Parinello and Rahman
\cite{PARR_FLUCM}. 
Here the information on the actual shape of the simulation volume, which is free to fluctuate in this ensemble,
is saved in the transformation matrix $\mathbf{h}$. One of the advantages of this implementation is, that from the fluctuations 
of the transformation matrix $\mathbf{h}$ the fluctuations of the strain tensor can be calculated directly. The strain tensor is related to the transformation matrix
via \cite{PARR_FLUCM}:
\begin{eqnarray}\nonumber
\mathbf{\epsilon}=\frac{1}{2}\left(\mathbf{h}_{0}^{T,-1}\mathbf{G}\mathbf{h}_{0}^{-1}-\mathbf{1}\right)
\end{eqnarray}
where $\mathbf{h}_{0}$ is the transformation matrix of the reference lattice and $\mathbf{G}=\mathbf{h}^{T}\mathbf{h}$ contains the information of the actual shape
of the simulation volume.
Table \ref{PRR_tabelle} shows a comparison of the elastic moduli for the harmonic, triangular lattice as they are expected for the chosen spring constant $f$ and the values as they are obtained from the
simulations in the $NpT$ ensemble by analyzing the fluctuations of the simulation volume.

Figure \ref{npt_ort} shows the strain-strain correlation functions in Fourier space as they are obtained from the simulations in the $NpT$ ensemble with periodic
boundary conditions. The eight-fold rotational symmetry in $\tilde{G}_{11}(\vec{k})$ is not resolved. The shear strain correlation functions show clearly a four-fold
rotational symmetry, as was expected from the analytic predictions. 
\begin{figure}
\begin{center}
\includegraphics[width=7.25cm]{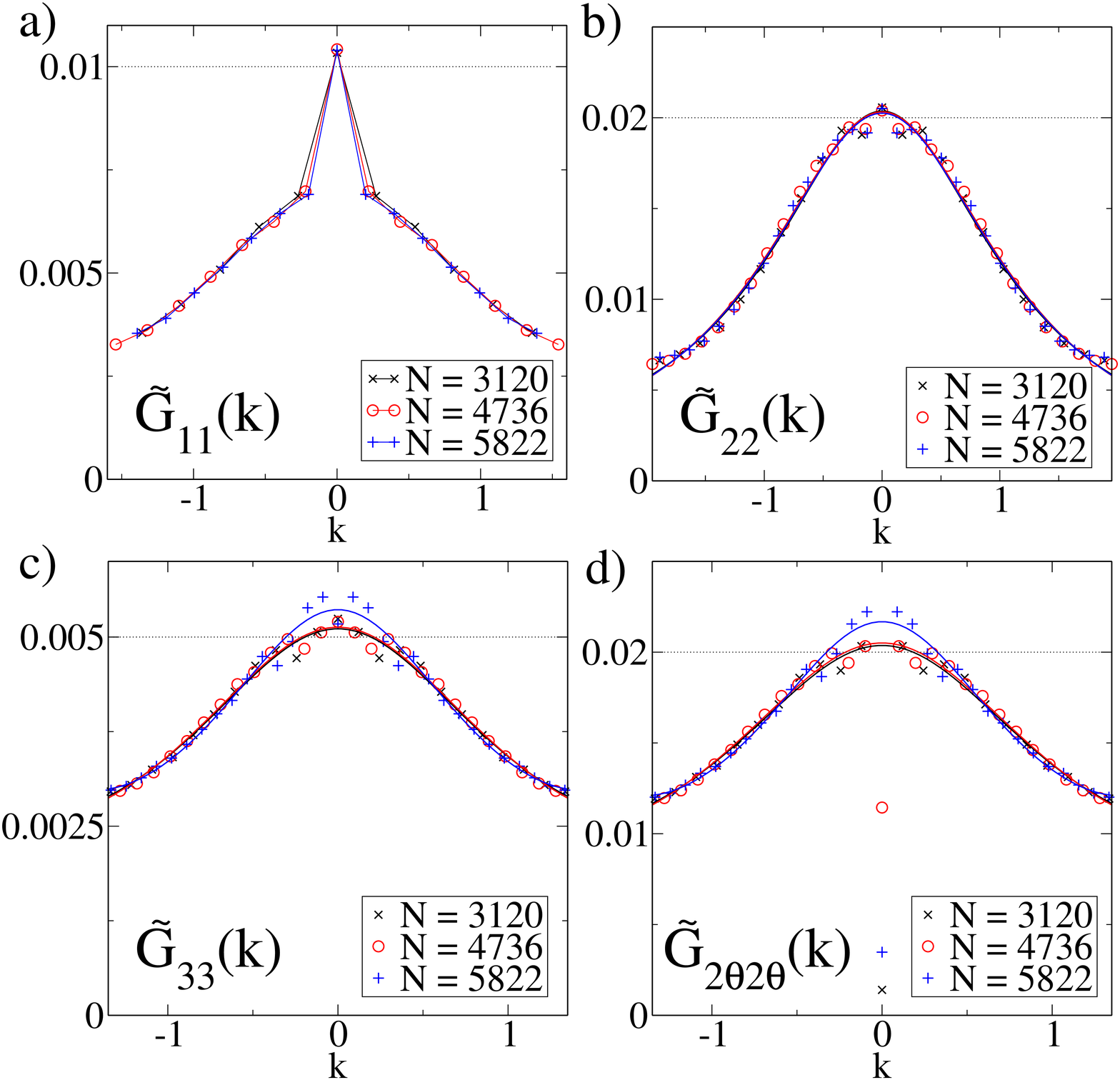}
\caption{(Color online) Cuts of the strain-strain correlation functions in the $NpT$ ensemble with periodic boundary conditions at zero hydrostatic 
pressure along specific directions in Fourier space.  a) $\tilde{G}_{11}(\vec{k})$ along $k_{y}=2k_{x}$, 
b) $\tilde{G}_{22}(\vec{k})$ along $k_{x}=k_{y}$, c) $\tilde{G}_{33}(\vec{k})$ and d) $\tilde{G}_{2\theta 2\theta}(\vec{k})$ along the
$k_{y}$-axis. For a comparison the results for different system sizes 
are displayed. The horizontal broken lines mark the value expected from
theory for the correlation functions at $\vec{k}=\vec{0}$.\label{nabh_g_npt}}
\end{center}
\end{figure}
Cuts along various directions in Fourier space of these functions
are plotted in figure \ref{nabh_g_npt} for the three system sizes. 
As these cuts in Fourier space show, there is no systematic dependence on the system size in the correlation functions.

The discontinuities in the correlation functions are visible in figure \ref{npt_ort} and \ref{nabh_g_npt}. Nevertheless the extreme discontinuity
one expects to observe in $\tilde{G}_{11}(\vec{k})$ from the analytic predictions is reduced to a factor of approximately
$1.5$ instead of $30$, as a comparison of the cuts in figure \ref{a_cuts2} and in figure \ref{nabh_g_npt} a) shows. This indicates that there might be excitations in the system that are not captured by the
assumption of purely affine strains. Along the cuts, for which $\tilde{G}_{22}(\vec{k})$ and $\tilde{G}_{33}(\vec{k})$ are continuous for $\vec{k}\rightarrow
\vec{0}$, fitting with a generalize Lorentzian profile yields the elastic constants and via the coefficients $c_{i}$ the elastic correlation lengths. For the system
with $N=3120$ we obtain the shear modulus as it is given in table \ref{PRR_tabelle} and the coefficients
$c_{2}=34.3$, $c_{2}'=-0.6$, $c_{3}= 114.1$ and $c_{3}'=-17.6$. So the elastic correlation lengths
$\xi_{el,i}\sim\sqrt{c_{i}}$ are approximately $6$ and $11$ lattice parameters respectively.

Figure \ref{nabh_g_npt} d) shows cuts along the coordinate axis of $\tilde{G}_{2\theta~2\theta}(\vec{k})$. 
In these simulations the system as a whole is not an embedded system and is not free to rotate. Thus we cannot obtain the elastic modulus 
directly from the value of  $\tilde{G}_{2\theta~2\theta}$ at the origin. This situation is different in a
solid, which is embedded in a larger volume, as will be discussed in section \ref{EMBED}.

\subsection{$NVT$ ensemble with periodic boundary conditions\label{NVTPBC}}
Below, we describe simulations in the $NVT$ ensemble with periodic boundary conditions at a reduced density of $\varrho^{*}=1.0$. 
Table \ref{SQR_tabelle}
lists the elastic constants calculated with the fluctuation method given by Squire et. al. \cite{SQUIRE}. These authors 
also give a formula for calculating the stress tensor.
An evaluation of the data yields $\sigma_{xy}=\sigma_{yx}=0.0$ and from the trace of the stress tensor $\beta
a^{2}p=-\frac{1}{2}(\sigma_{xx}+\sigma_{yy})\approx 0.1$. So we verify that the simulations represent a solid at approximately zero hydrostatic pressure.
\begin{table}[t!]
\begin{center}
\begin{tabular}{|l|r|r|r|}
\hline
~& $a_{1}$& $a_{2}$& $a_{3}$\\
\hline
~&~&~&~\\
calculated from $f$ &$100~/\beta a^{2}$&$50~/\beta a^{2}$&$200~/\beta a^{2}$\\
\hline
~&~&~&~\\
using Squire et al. \cite{SQUIRE}&$97.4~/\beta a^{2}$&$48.3~/\beta a^{2}$&$194.8~/\beta a^{2}$\\
\hline
~&~&~&~\\
from fits of $\tilde{G}(\vec{k}\neq \vec{0})$&-&$48.1~/\beta a^{2}$&$198.2~/\beta a^{2}$\\
\hline
\end{tabular}
\end{center}
\caption{A comparison of the elastic constants as calculated from the spring constant $f$ and as obtained by use of various methods from data 
of a Monte Carlo simulation of a harmonic triangular lattice in the $NVT$ ensemble with periodic boundary conditions 
with $N=3120$ particles and spring constant $\beta a^{2} f= 200/\sqrt{3}$.\label{SQR_tabelle}}
\end{table}
\begin{figure}
\begin{center}
\includegraphics[width=7.25cm]{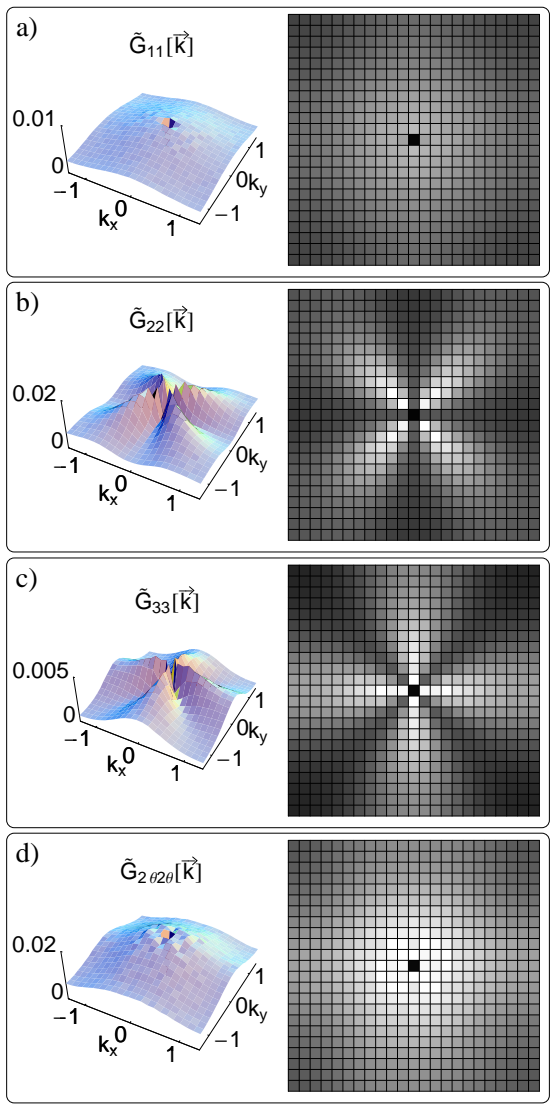}
\caption{(Color online) Strain-strain correlation functions of a harmonic triangular lattice as obtained from Monte
Carlo simulations in the $NVT$ ensemble with periodic boundary conditions. Results for s system with  $N=3120$ particles and a
$\beta a^{2} f= 200/\sqrt{3}$ are shown. For each function a surface plot and next to it a density plot
are displayed. In the density plot the maxima are white, minima are black.\label{nvt_g_ort} }
\end{center}
\end{figure}
\begin{figure}
\begin{center}
\includegraphics[width=7.25cm]{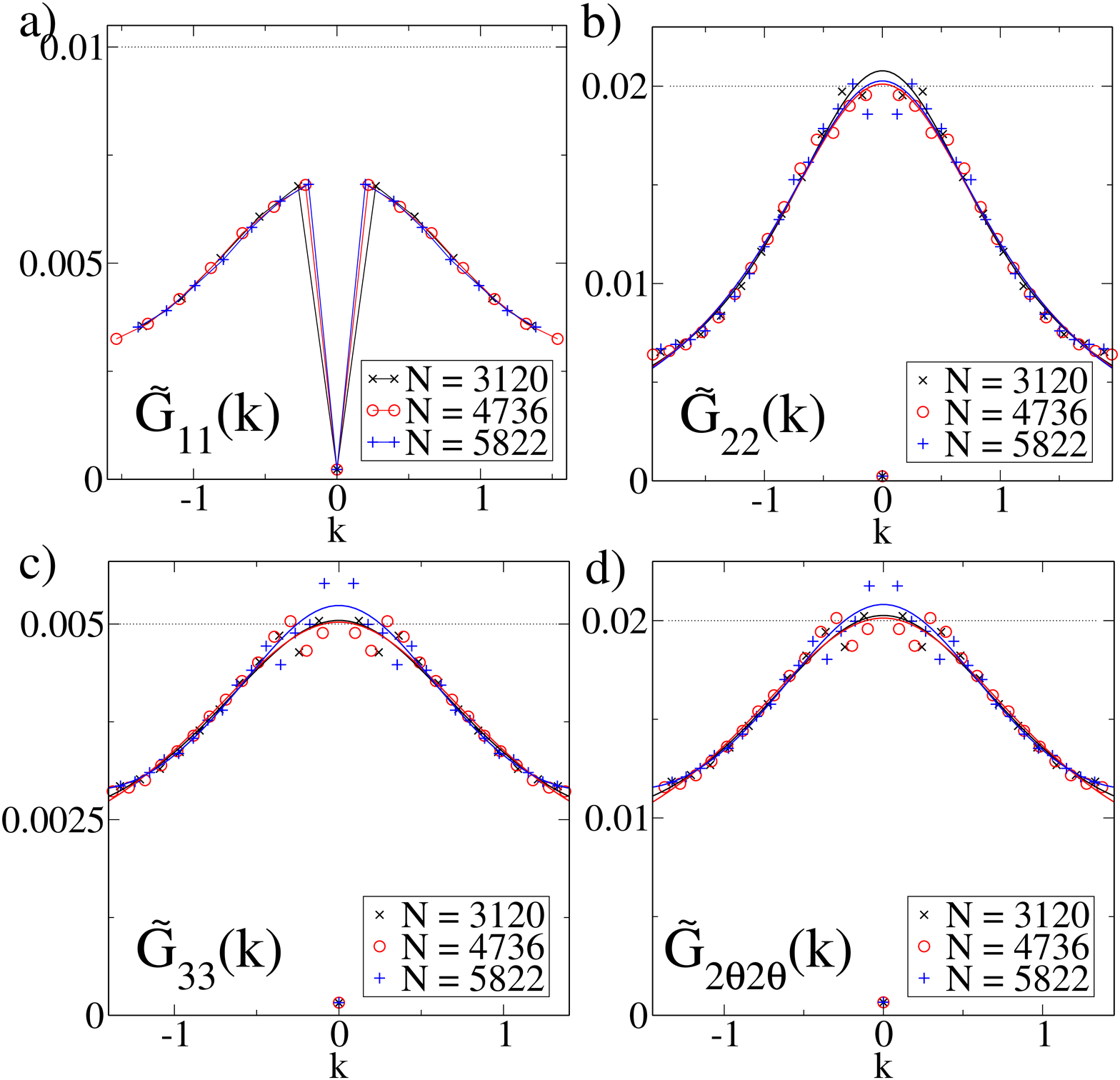}
\caption{(Color online) Cuts of the strain-strain correlation functions in the $NVT$ ensemble with periodic boundary conditions 
 along specific directions in Fourier space.  a) $\tilde{G}_{11}(\vec{k})$ along $k_{y}=2k_{x}$, 
b) $\tilde{G}_{22}(\vec{k})$ along $k_{x}=k_{y}$, c) $\tilde{G}_{33}(\vec{k})$ and d) $\tilde{G}_{2\theta 2\theta}(\vec{k})$along 
the $k_{y}$-axis. For a comparison the results for different system sizes are displayed. The horizontal broken lines mark the value expected from
theory for the correlation functions at $\vec{k}=\vec{0}$.\label{nabh_g_nvt}}
\end{center}
\end{figure}

In this ensemble the $\vec{k}=\vec{0}$ values of the correlation functions cannot be used to calculate the elastic
constants directly. We are simulating an undeformed state of the crystal, thus the integral over the fluctuations of the
strains over the complete simulation volume tends to zero in this ensemble. Therefore only fits along the
directions, for which the correlation functions are continuous for $\vec{k}\rightarrow\vec{0}$, give access to the elastic
constants in this ensemble. As  $\tilde{G}_{11}(\vec{k})$ has no such direction, the bulk modulus cannot be obtained in this way. 
From these considerations one expects the strain-strain correlation functions in the $NVT$ ensemble to differ from those in the $NpT$ ensemble for small absolute
values of $\vec{k}$. Figure \ref{nvt_g_ort} shows surface plots and density plots of the strain-strain correlation functions in
Fourier space. As in section \ref{NPTPBC} the anisotropies are recovered well, except that the eightfold rotational symmetry of
$\tilde{G}_{11}(\vec{k})$ is not resolved. The expected discontinuous jump to $\tilde{G}_{ii}(\vec{k}=\vec{0})=0$ ($i=1$, $2$, $3$) is clearly visible. Besides
this, the correlation functions coincide with those obtained in the $NpT$ ensemble, as one can see by comparing figure 
\ref{nabh_g_npt} and \ref{nabh_g_nvt}, showing
the same cuts in Fourier space for the various correlation functions. The elastic constants $a_{2}$ and $a_{3}$, as they are obtained 
from fitting the strain-strain correlation functions, are listed in table \ref{SQR_tabelle}.
They fall within $3-4\%$ of the theoretical values and have thus the same accuracy as the values obtained via Squire's fluctuation formulae \cite{SQUIRE}. In addition the 
elastic correlation lengths could be obtained from the coefficients: $c_{2}=37.4$, $c_{2}'=-1.3$, $c_{3}=118.2$ and $c_{3}'=18.8$. So
consistent to the results obtained from the simulations in the $NpT$ ensemble $\xi_{el,i}\sim 6$ and $11$ lattice parameters respectively. Fitting e.g.
$\tilde{G}_{22}(\vec{k}\neq \vec{0})$ along the direction $k_{y}=2k_{x}$ allows the determination of the coefficients $c_{1}=54.3$ and
$c_{1}'=-86.1$. Thus correlations of volume fluctuations decay over approximately 7 lattice parameters. 

The harmonic, triangular lattice in the $NVT$ ensemble was also analyzed with \emph{open} boundary conditions. The results were discussed in detail in
\cite{KFRANZ2}. 

\subsection{The influence of hydrostatic pressure\label{PRESS}}
\begin{figure}
\begin{center}
\includegraphics[angle=-0,width=7.25cm]{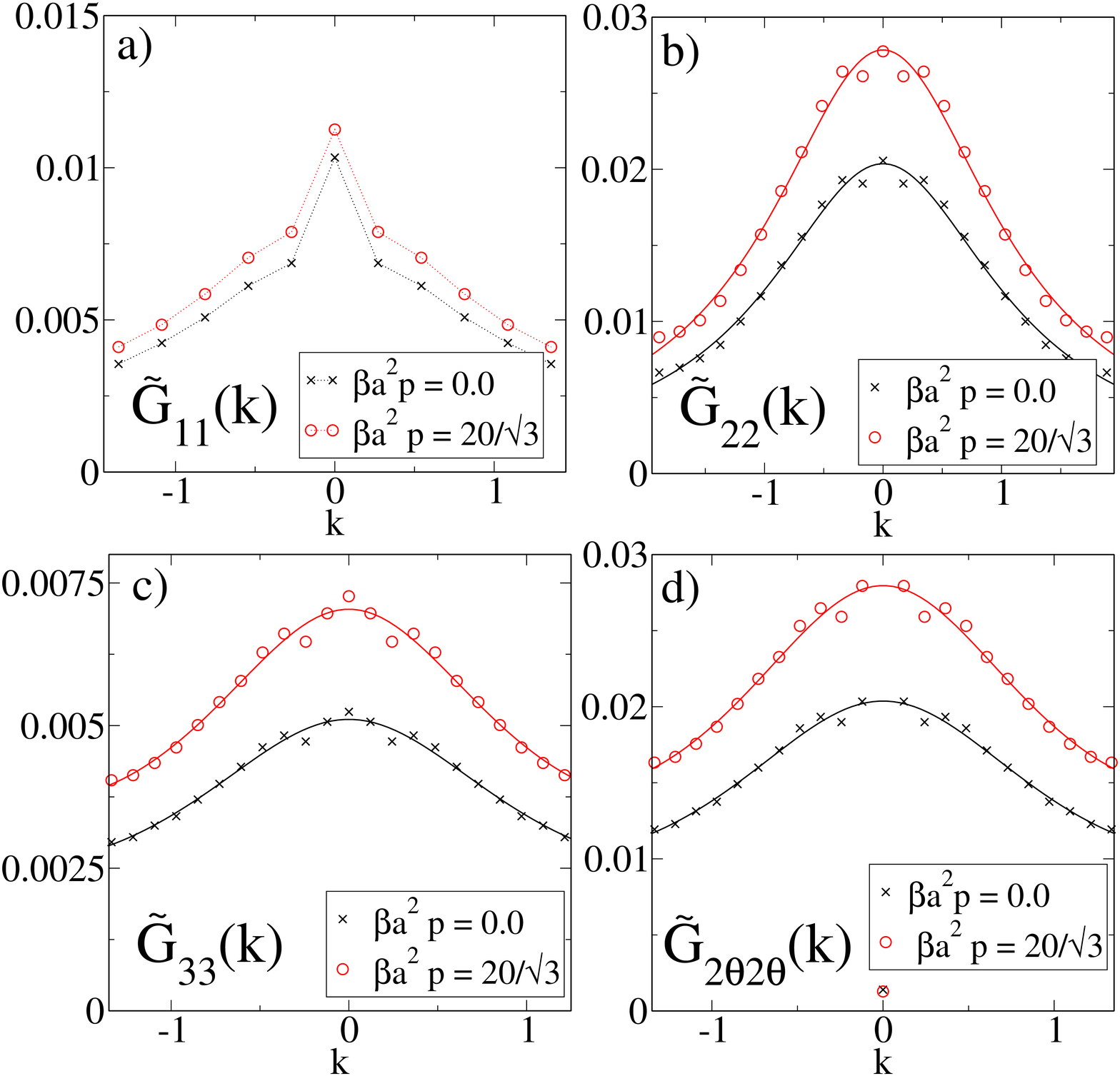}
\caption{(Color online) Cuts of the strain-strain correlation functions in the $NpT$ ensemble with periodic boundary conditions 
 along specific directions in Fourier space. Shown are the correlation functions obtained in a system without (crosses) 
 external pressure and a system  with $\beta a^{2} p =20/\sqrt{3}$ (circles). Both systems have $N=3120$ and a spring constant
  $\beta a^{2} f=200/\sqrt{3}$. a) $\tilde{G}_{11}(\vec{k})$ along $k_{y}=2k_{x}$, b) $\tilde{G}_{22}(\vec{k})$ along $k_{x}=k_{y}$, 
 c) $\tilde{G}_{33}(\vec{k})$ and d) $\tilde{G}_{2\theta 2\theta}(\vec{k})$ along the $k_{y}$-axis.\label{cut_druckvg}}
\end{center}
\end{figure}

How does an external, hydrostatic pressure - i.e. $\sigma_{xy}=\sigma_{yx}=0$  and
$\sigma_{xx}=\sigma_{yy}=-p$ - influence the strain-strain correlation functions? Simulations of a harmonic,
triangular lattice with spring constant $\beta a^{2} f=200/\sqrt{3}$ subjected to an external, hydrostatic 
pressure $\beta a^{2} p =20/\sqrt{3}$ show, that the shape of the correlation functions is not affected. The $NpT$ ensemble was
chosen for this study. Strains were calculated with respect to the average lattice positions, that is the compressed lattice. The
lattice parameter of this reference lattice $a'=1.010465$ is smaller than the lattice parameter $a=(2/\sqrt{3})^{1/2}$ in the 
zero-pressure simulations. Therefore for a comparison with the theoretical values, which were given in units of $\beta a^{2}$, 
the $a_{i}$ as they are obtained e.g. from $\tilde{G}_{ii}(\vec{k}=\vec{0})$ must be rescaled to these units. Simulations were run for a
system with $N=3120$ particles. For a direct comparison of the correlation functions in systems with and without a hydrostatic
pressure cuts in Fourier space of the $\tilde{G}_{ii}(\vec{k})$ ($i=1,2,3,2\theta$) are shown in figure \ref{cut_druckvg}.
\begin{table}[b!]
\begin{center}
\begin{tabular}{|l|r|r|r|}
\hline
~&~&~&~\\
calculated&~&$\mu=a_{2}+p$&$4\mu=a_{3}+4p$\\
from $f$ &$100~/\beta a^{2}$&$=50~/\beta a^{2}$&$=200~/\beta a^{2}$\\
\hline
~&~&~&~\\
from&~&~&~\\
$\tilde{G}(\vec{k}=\vec{0})$&$88.8~/\beta a'^{2}$&$a_{2}=36.0~/\beta a'^{2}$&$a_{3}=137.6~/\beta a'^{2}$\\
\hline
~&~&~&~\\
rescaled&~&~&~\\
values&$100.4~/\beta a^{2}$&$a_{2}= 40.7~/\beta a^{2}$&$a_{3}=155.6~/\beta a^{2}$\\
~&$~$&$\rightarrow \mu=52.2~/\beta a^{2}$&$\rightarrow 4\mu=201.8~/\beta a^{2}$\\
\hline
\end{tabular}
\end{center}
\caption{The elastic moduli of the harmonic system calculated from the spring constant $\beta a^{2} f=200/\sqrt{3}$ in comparison 
to simulation results. Listed are the elastic constants $a_{i}$ as obtained from $\tilde{G}_{ii}(\vec{k}=\vec{0})$ for a simulation 
in the $NpT$ ensemble with periodic boundary conditions of a harmonic, triangular lattice with $\beta a^{2} p =20/\sqrt{3}$. 
For comparison with the theoretical values the $a_{i}$ obtained in units of the lattice parameter $a'$ of the compressed reference
lattice have to be rescaled to  the lattice parameter $a$ of the zero-pressure reference lattice and the relation between the stiffness tensor
$B_{ijkl}$ and the tensor of elastic constants $C_{ijkl}$ in a system with $\sigma_{ij}\neq 0$ need to be considered. \label{tabel_druck}}
\end{table}
These show clearly the shift in the absolute values and the persistence of their shape. When relating the parameter $a_{i}$ to the elastic moduli of
the system, one has to recall, that the strain-strain correlations are related to the stiffness tensor $B_{ijkl}$, which is defined via the
stress-strain relations. These relate the variation of stress to the variation of strain, to first order in the strains, for the case, that an arbitrary
initial configuration $\{\vec{R}\}$ is transformed to a final configuration $\{\vec{r}\}$ by an applied uniform stress. 
Thus $B_{ijkl}(\{\vec{R}\})\equiv \left(\partial \sigma_{ij}(\{\vec{r}\})/\partial
\epsilon_{kl}\right)_{\{\vec{R}\}}=\frac{1}{2}\left(\sigma_{il}\delta_{jk}+\sigma_{jl}\delta_{ik}+\sigma_{ik}\delta_{jl}+\sigma_{jk}\delta_{il}-2\sigma_{ij}\delta_{kl}\right)+C_{ijkl}$
\cite{WALLA}. The stiffness tensor explicitly depends on the applied stress. Only for the
case that $\sigma_{ij}=0$ is it equivalent to the tensor of the elastic constants $C_{ijkl}$. For the calculation of 
the elastic moduli the following combinations are needed:
\begin{eqnarray}\nonumber
\frac{1}{2}\left(B_{xxxx}+B_{xxyy}\right) &=&\frac{1}{2}\left(C_{xxxx}+C_{xxyy}\right)-\frac{1}{2}\left(p-p\right)\\\nonumber
 &=& K\\\nonumber
\frac{1}{2}\left(B_{xxxx}-B_{xxyy}\right) &=&\frac{1}{2}\left(C_{xxxx}-C_{xxyy}\right)-\frac{1}{2}\left(p+p\right)\\\nonumber 
 &=&\mu - p\\\nonumber
B_{xyxy} &=&C_{xyxy}-p =\mu - p
\end{eqnarray}
Thus the bulk modulus can be directly obtained from $a_{1}$, while from $\tilde{G}_{22}(\vec{k})$ one extracts $a_{2}=\mu - p$ and 
from $\tilde{G}_{33}(\vec{k})$ one obtains $a_{3}=4(\mu - p)$. The values of the elastic moduli calculated according to this
scheme are given in table \ref{tabel_druck}. They lie within $4.4\%$ of the theoretical values. 

\subsection{Effects of an embedding medium and finite-size\label{EMBED}}
\begin{figure}
\begin{center}
\includegraphics[width=7.25cm]{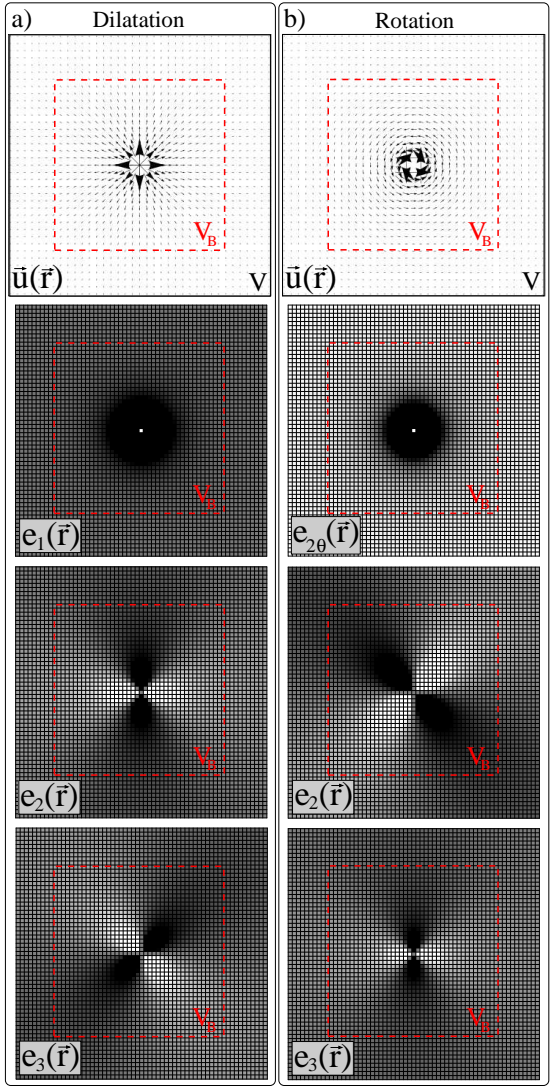}
\caption{(Color online) A schematic drawing of the displacement field $\vec{u}(\vec{r})$ as it results from a given perturbation at the origin: 
a) a dilatation and b) a rotation a the disk at the origin. If the analyzed volume $V_{B}$ (broken red line) does not coincide with the volume
of the complete system $V$ (black line), the energy needed for the displacements in the embedding medium ($V-V_{B}$) must be taken into
account in the interpretation of the correlation functions.\label{embedd_scheme}}
\end{center}
\end{figure}
Often the strain-strain correlations cannot be evaluated over the complete crystal. In experiments as for example a two dimensional colloidal 
crystal \cite{ZAHN,KFRANZ2} configurational data is taken via video microscopy. Here the area accessible to the video camera is far smaller than the 
complete sample size. In these cases only a sub-system embedded in a larger continuum is analyzed. As Zahn et al. \cite{ZAHN} noted the presence of 
an infinite, embedding medium alters the relation of the strain fluctuations to the elastic moduli. 

The strain-strain correlation functions are the 
response functions to a strain perturbation at the origin. Figure \ref{embedd_scheme} shows schematically the resulting displacement field and strain
fields for the cases that this perturbation is a) a dilatation and b) a rotation. The connection between the strain correlations and the elastic moduli
was derived under the assumption, that the considered functional of the free energy accounts for the free energy of the complete system (equipartition
theorem). For the case that the volume $V_{B}$ over which the strain-strain correlation function are calculated is not equal to the complete system
volume $V$ this assumption is not fulfilled any more. As can be seen in the schematic plots of the strain fields in figure \ref{embedd_scheme} the energy
related to the resulting strain field outside $V_{B}$ cannot be neglected for $V_{B}\neq V$. 
  
Following the argument by Zahn et al. \cite{ZAHN}, but considering a \emph{finite} embedding continuum, we show that the influence of the 
surrounding medium on the strain fluctuations within the analyzed volume $V_{B}$ depends on the relative size of $V_{B}$ in comparison to the complete system volume $V$,
i.e. the ratio of $V_{B}/V$. For the derivation of the formulae we consider first a homogeneous dilatation of a disk $V_{B}=\pi R_{B}^{2}$ in a
surrounding medium of volume $V=\pi R^{2}$ and second a pure shear, which can be realized by a rotation by an angle $\theta$ of the
disk with volume $V_{B}$. For these considerations we work in polar coordinates, where we have 
$e_{1}=(\epsilon_{rr}+\epsilon_{\varphi\varphi})$, $e_{2}=(\epsilon_{rr}-\epsilon_{\varphi\varphi})$ and
$e_{3}=\epsilon_{r\varphi}$. In both cases considered here it is assumed that the displacement field on the boundary of the
complete system is given by $\vec{u}(\vec{r}=\vec{R}_{boundary})=\vec{0}$.  

\subsubsection{Homogeneous dilatation\label{HOMDIL}} 
An isotropic expansion of a disk embedded in a finite medium is given by: $R_{B}\rightarrow R_{B}+\Delta r$. The resulting displacement field in polar coordinates
is given by:
\begin{eqnarray}\nonumber
u_{\varphi} = 0~~\textrm{,}~~~~u_{r}=\Bigg\{{\Delta r\frac{r}{R_{B}}\qquad \textrm{for}\qquad r<R_{B}\atop{\Delta r\frac{R_{B}}{r}\qquad \textrm{for}\qquad r>R_{B}}}
\end{eqnarray}
From this it is straight forward to calculate the resulting strain field and consequently the Free Energy density 
$f=\frac{1}{2}\left[K(\epsilon_{rr}+\epsilon_{\varphi\varphi})^{2}+\mu\left((\epsilon_{rr}-\epsilon_{\varphi\varphi})^{2}+4\epsilon_{r\varphi}^{2}\right)\right]$
of the system under load. Thus the total energy needed for such an expansion in a finite system of volume $V=\pi R^{2}$ is given by $E=\int_{0}^{2\pi}\int_{0}^{R}rd\varphi dr~f=\left(V_{B}/2\right)\left(\Delta
 V_{B}/V_{B}\right)^{2}\left[K+\mu\left(1-\left(V_{B}/V\right)\right)\right]$. 
For such a system equipartition tells us thus, that the strain fluctuations are no longer set by the bulk
modulus of the system, but aquire in the embedded system a term dependent on the shear modulus {\emph{and}} on the
ratio $V_{B}/V$:
\begin{eqnarray}\nonumber
\frac{k_{B}T}{V_{B}}\langle e_{1}^{2}\rangle=\frac{1}{K+\mu\left(1-\left(\frac{V_{B}}{V}\right)\right)}
\end{eqnarray}
Therefore the strain-strain correlation function $G_{11}(\vec{r})$ no longer provides access to the bulk modulus,
but to a $V_{B}/V$-dependent combination of bulk and shear modulus. 
\subsubsection{Pure shear\label{PUSHE} } 
A rotation of the disk as a rigid body within the embedding medium by an infinitesimal angle $\Delta\varphi$ changes a given 
orientation $\varphi$ to $\varphi+\Delta\varphi$. 
The resulting displacement field is given by:
\begin{eqnarray}\nonumber
u_{r}=0~~\textrm{,}~~~~u_{\varphi}=\Bigg\{{0\qquad\qquad\quad\textrm{for}\qquad r<R_{B}\atop{\Delta \varphi\frac{R_{B}^{2}}{r}\qquad\quad\textrm{for}\qquad r>R_{B}}}\nonumber
\end{eqnarray}
From the corresponding strain field the Free Energy density can be determined and integration over the complete system yields the energy required for such a
rotation: $E=\mu(2\Delta\varphi)^{2}V_{B}\left(1-\left(V_{B}/V\right)\right)/2$. In case of infinitesimal rotation angles $\Delta \varphi$ this angle can be identified with the anti-symmetric part of the strain tensor
$\theta=\left(\partial u_{y}/\partial x-\partial
u_{x}/\partial y\right)/2$. Equipartition relates the fluctuations in $e_{2\theta}$ to the shear modulus $\mu$:
\begin{eqnarray}\nonumber
\frac{k_{B}T}{V_{B}}\langle e_{2\theta}^{2}\rangle=\frac{1}{\mu\left(1-\left(\frac{V_{B}}{V}\right)\right)}\nonumber
\end{eqnarray}
This relation depends also on the ratio $V_{B}/V$, as the energy required for the rotation of a disk, which is not embedded in a surrounding
medium, tends to zero.
The analysis of the strain variable $2\theta$ offers thus an independent, direct route to the determination of the
shear modulus in an embedded system. 

These considerations show that in order to obtain accurate elastic moduli from the analysis of the strain fluctuations the
relative size of the analyzed system to the complete, finite system should to be known. Nevertheless for the case of the colloidal
crystal \cite{KFRANZ2} the situation is close to the limiting case of  $\frac{V_{B}}{V}\rightarrow 0$. Here the influence of the
surrounding medium on the analyzed system is dominant. The strain variables $e_{1}$ and $e_{2\theta}$ can be used to
extract the elastic moduli. Thus the two-dimensional colloidal crystal, as discussed in detail in
\cite{KFRANZ2,KFRANZ3} is an example for a completely embedded system. In contrast to this in simulations the complete system can be analyzed, which corresponds to
the limiting case of $\frac{V_{B}}{V}\rightarrow 1$. For this case each of the strain-strain correlation functions
of the strain variables $e_{1}$, $e_{2}$ and $e_{3}$ give directly access to the corresponding elastic moduli $a_{i}$. The effect of the embedding medium can
be visualized by looking at a statistical sum rule, as will be discussed in the next paragraph.

\subsubsection{Analysis of a statistical sum rule\label{SUMR}}
The sum rule for the generalized susceptibility provides another way of extracting the elastic moduli from 
the strain-strain correlation functions. The coarse-grained system represents a homogeneous continuum and is thus translationally invariant. 
For such 
systems the susceptibilities are directly related to the correlation functions. In Fourier space this 
reads $\tilde{\chi}_{T}(\vec{k})=\beta\tilde{G}(\vec{k})$. From this the static susceptibility sum rule follows \cite{GOLDF}:
\begin{eqnarray}\nonumber
\chi_{T}= \lim_{\vec{k}\to 0} \tilde{\chi}_{T}(\vec{k})=\beta\tilde{G}(\vec{k})|_{\vec{k}=\vec{0}}= \beta\int d\vec{r}~G(\vec{r})
\end{eqnarray}
Thus an integration of the correlation functions in real space yields directly the elements of the compliance tensor $S_{ijkl}$, which 
correspond to the generalized susceptibilities. The compliance tensor is the inverse of the stiffness tensor
$B_{ijkl}$. In the case that \emph{no} external stresses act on the system this is equivalent to the tensor of the elastic constants $C_{ijkl}$ 
\cite{WALLA}. The $S_{ijkl}$ obtained from such an analysis depend on the integration volume $V_{B}$. Thus in order to obtain systems-size independent
values an additional finite-size scaling analysis should be employed. 
\begin{figure}
\begin{center}
\includegraphics[angle=0,width=7.25cm]{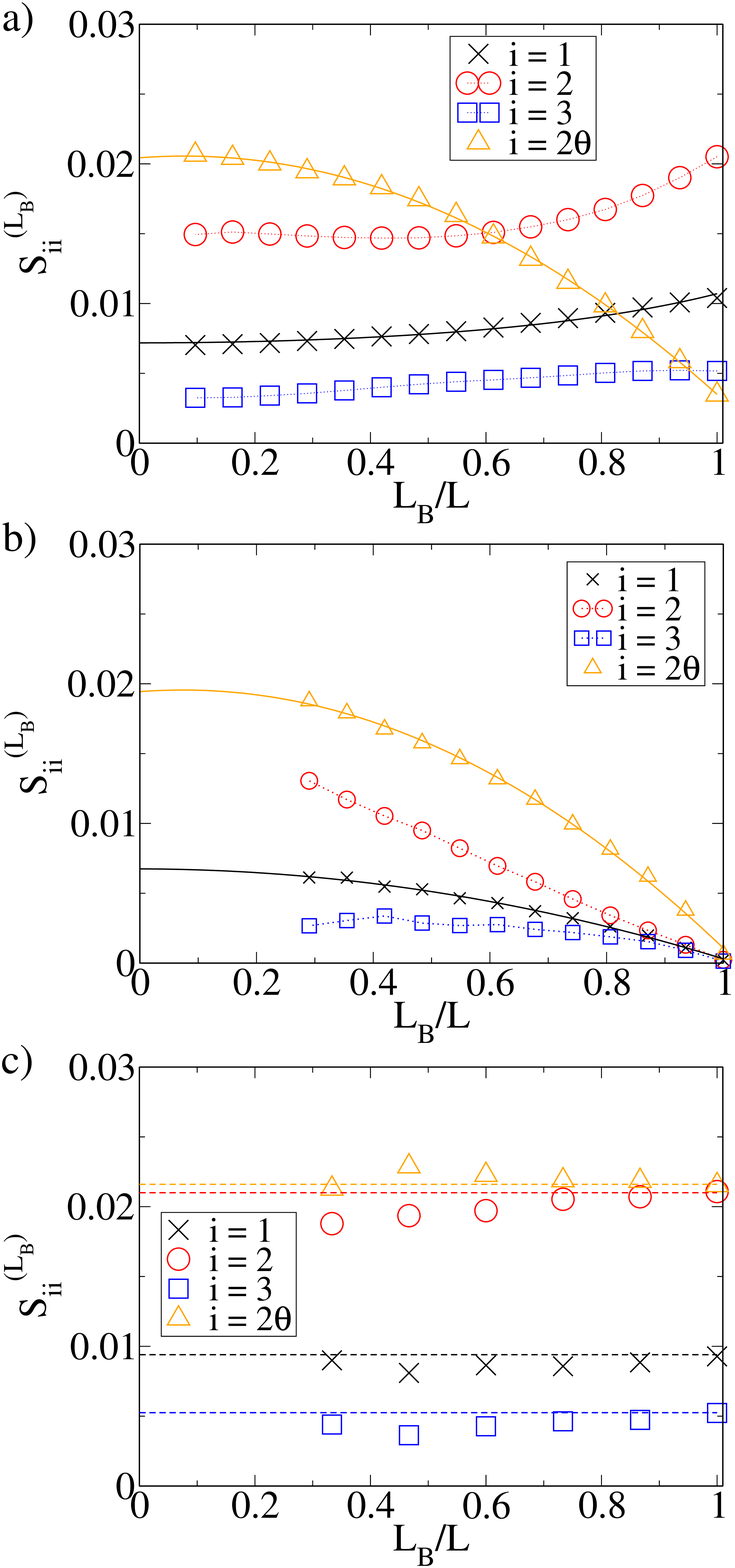}
\caption{(Color online) Compliances $S_{ii}$ as they are obtained from the sum rule as a function of the ratio of analyzed volume $V_{B}=L_{B}^{2}$ to the complete simulation
volume $V=L^{2}$. Shown are the results of Monte Carlo simulations of a harmonic triangular lattice with $N=5822$ particles and spring constant $\beta
a^{2} f= 200/\sqrt{3}$. Lines are a fit to the
data with the formulae given in the text, while the dotted lines are a guide for the eyes. a) $NpT$ ensemble with periodic boundary conditions. b) $NVT$
ensemble with periodic boundary conditions. c) $NVT$ ensemble with open boundary conditions. Here dashed lines show the value for $L_{B}/L=1$ as a comparison. \label{vel_npt}}
\end{center}
\end{figure}

Figure \ref{vel_npt} 
shows the compliances $S_{ii}$ ($i=1$, $2$, $3$ and $2\theta$) as a function of the ratio of the integration volume $V_{B}$ to the
 complete simulation volume $V$, i.e. $V_{B}/V=L_{B}/L$, as they are obtained from the simulation data of the harmonic triangular crystal at zero external
 pressure in the $NpT$ and $NVT$ ensemble 
 with different boundary conditions. A comparison shows directly how the choice of ensemble and the choice of boundary conditions influences the results. These are
 so called explicit and implicit  finite-size effects \cite{ROMAN2}. 
 In addition this analysis visualizes the effects of the embedding medium on the compliances $S_{ii}$. Figure \ref{vel_npt} a) shows the results 
from simulations in the $NpT$ ensemble with periodic boundary conditions. The complete system contains $N=5822$
particles, that are connected via springs of spring constant  $\beta a^{2} f= 200/\sqrt{3}$. The strain-strain correlation functions are directly related to 
the elastic moduli in this ensemble, due to the fact that the volume itself fluctuates. Thus one can obtain the elastic moduli directly from the
$S_{ii}$ at $L_{B}/L=1$. 
For the considered system we find: $S_{11}^{(L)}=0.01025$, $S_{22}^{(L)}=0.02050$, $S_{33}^{(L)}=0.00512$ and $S_{2\theta 2\theta}^{(L)}\rightarrow 0$, resulting in $a_{1}=97.5$,
$a_{2}=48.7$ and $a_{3}=195.2$. These values lie within $2.5\%$ of the expected values. The accuracy of this approach compares to
that of the methods for the calculation of the elastic moduli discussed before. 
From the considerations in \ref{HOMDIL} one expects the following functional dependence of $S_{11}$ on $L_{B}/L$:
\begin{eqnarray}\nonumber
S^{(L_{B})}_{11}&=&V_{B}\langle e_{1}^{2}\rangle/(k_{B}T)=\left[K+\mu\left(1-\left(L_{B}^{2}/L^{2}\right)\right)\right]^{-1}\label{erw_Km}\nonumber
\end{eqnarray}
The black solid line in figure \ref{vel_npt} a) is a fit with this equation to the data (crosses). From the fit parameters the following elastic moduli are
extracted: $a_{1}=K=93.4$ and $\mu=a_{2}=45.8$. Figure \ref{vel_npt} a) shows clearly the increasing impact the surrounding medium has on $S_{11}$ as $L_{B}/L$ diminishes. 
In the limit $V_{B}/V\rightarrow 0$ it 
yields the sum of the elastic moduli $K$ and $\mu$. It is apparent from figure \ref{vel_npt} a), that as soon as $V_{B}/V<1$, the compliances $S_{22}$ and $S_{33}$ cannot be directly related to the shear modulus any more.

The considerations in \ref{PUSHE} suggest, that the compliance $S_{2\theta 2\theta}$ should diverge as 
$V_{B}/V\rightarrow 1$. This relates to the fact, that the energy needed for rotating an embedded disk goes to zero as the embedding material is removed. 
This divergence cannot be seen in the simulation data, as in simulations with periodic boundary condition the system as a whole cannot rotate.  
The fact, that there is no divergence of $S_{2\theta 2\theta}$ for $V_{B}/V\rightarrow 1$, 
is therefore an implicit finite-size effect. In order to extract the shear
modulus from the compliance $S_{2\theta 2\theta}$ a polynomial in $(L_{B}/L)$ was fitted to the data (open triangles). From the
limit $V_{B}/V\rightarrow 0$ the shear modulus is extracted: $\mu=48.9$. 

In the $NVT$ ensemble with periodic boundary conditions the compliances as a function of $L_{B}/L$ exhibit a different
dependence on $L_{B}/L$, as figure \ref{vel_npt} b) shows. An unstrained state of the triangular lattice is analyzed in these simulations, therefore the integral of the correlation
functions over the complete system goes to zero and gives no access to the elastic moduli. This is an explicite finite-size effect. Nevertheless from 
the limit $V_{B}/V\rightarrow 0$ one can extract the elastic moduli as in the case of the simulations in the $NpT$ ensemble 
with periodic boundary conditions from the compliances $S_{11}$ (crosses) and $S_{2\theta 2\theta}$ (open triangles). Fits with a polynomial 
in $(L_{B}/L)$ are plotted as solid lines in figure \ref{vel_npt} b). From the case of maximum embedding one extracts $a_{1}+a_{2}=K+\mu=148.3$ from $S_{11}$ and $a_{2}=51.5$ from $S_{2\theta 2\theta}$ in figure \ref{vel_npt} b). These values compare
to the values obtained by different methods as they are given in table \ref{SQR_tabelle}. 

The compliances $S_{ii}^{L_{B}}$ shown in figure \ref{vel_npt} c) are obtained from data of simulations in the $NVT$ ensemble with open boundary 
conditions as they were presented in \cite{KFRANZ2}. These show in contrast no systematic dependence 
on $V_{B}/V$. The maximum analyzed volume, which will for this case be denoted by $V=L^{2}$, is approximately one fourth of the complete system volume. 
Averaging over the positions of origin, as it is done in the calculation of the strain-strain correlation functions in the system with open boundaries, results 
in an averaging over sub-systems with partial to complete embedding. For this type of 
averaging the $\vec{k}=\vec{0}$ values of \emph{all} considered strain-strain correlation functions give access 
to the elastic moduli of the system \cite{KFRANZ2}. The effect of this type of averaging shows up most prominently in the fact that
$S_{2\theta 2\theta}$ does not tend to zero for $L_{B}/L\rightarrow 1$, but approaches the value of $S_{22}$. Extracting the elastic moduli 
from the compliances for $L_{B}/L=1$ in figure \ref{vel_npt} c) yields  $S^{(L)}_{11}= 0.00940 \rightarrow a_{1}= 106.9$, $S^{(L)}_{22}= 0.02100 \rightarrow a_{2}= 47.6$, $S^{(L)}_{33}=
0.00525 \rightarrow a_{3}= 190.4$ and $S^{(L)}_{2\theta 2\theta}=0.02160 \rightarrow  a_{2}= 46.4$. The accuracy of these values is the
same as in \cite{KFRANZ2}. The deviation from the theoretical values is larger in this case, as the finite system with open boundary conditions is influenced in its elastic properties by the missing,
stabilizing bonds for particles at the surfaces.

\section{The statistics of non-affine fluctuations\label{NAFSTAT}}
\begin{figure}[b]
\begin{center}
\includegraphics[angle=0,width=7.25cm]{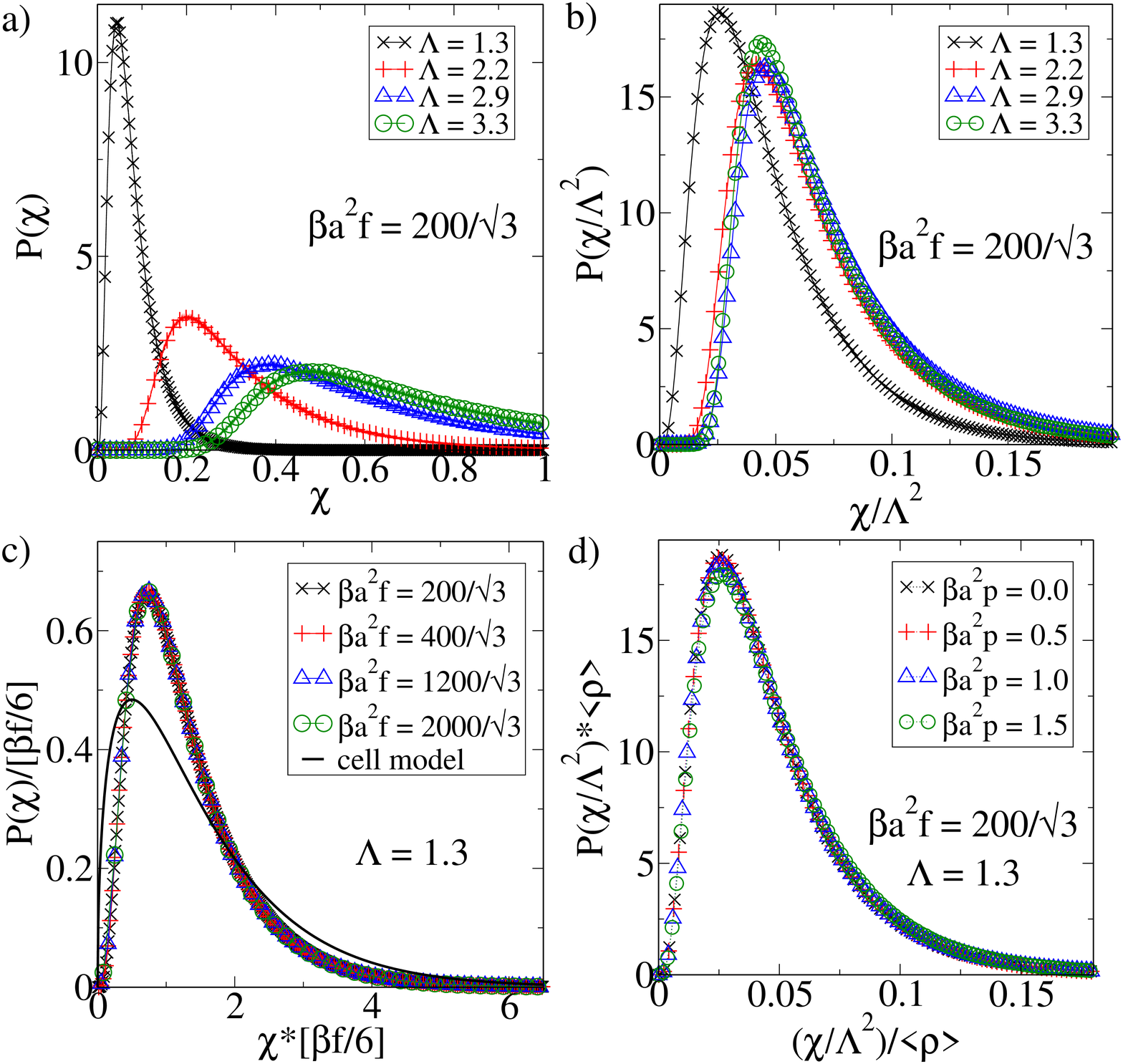}
\caption{(Color online) Plots of the probability distribution of the non-affinity parameter $\chi$ from simulations in the $NpT$ ensemble with periodic boundary
conditions and $N = 3120$ particles. (a) Plot of the probability distribution of $P(\chi)$ vs $\chi$ for various coarse-graining length $\Lambda$.
(b) The probability distribution of $\chi/\Lambda$ shows a data collapse for $\Lambda\geq 2.2$. (c) Data for various spring constants $f$ collapse
onto each other. The prediction from a simple cell model (black line) is shown for comparison. (d) Data from simulations at various hydrostatic
pressure $p$ scale with the resulting average density $\langle\varrho\rangle$.\label{CHISCA}}
\end{center}
\end{figure}
The coarse-graining process described in section \ref{SIMUS} projects the configurations generated by our microscopic 
Hamiltonian onto strain fields which are smooth over distances larger than the coarse-graining length $\Lambda$. It 
also generates a conjugate noise \cite{CHAIKI} which represents those fluctuations which cannot be captured during 
coarse-graining. This is easily understood once it is realized that, coarse-graining retains only that part of the 
particle displacements $\vec{u} = \vec{r}-\vec{R}$ in a configuration which can be obtained from the reference 
lattice $\vec{R}$ by an affine transformation: $\vec{r}=(\mathbf{1}+\epsilon)\vec{R}$. An affine transformation 
constrains all parallel lines in the reference lattice to remain parallel, which is clearly impossible to satisfy 
for an arbitrary configuration coarse-grained over volumes larger than an unit cell. Indeed, the quantity $\chi$ as 
defined in Eq. (\ref{CHIDEF}) has the dimension of Length$^2$ and scales as $\Lambda^2$. This may be seen by comparing 
figure \ref{CHISCA} (a) and (b). Figure \ref{CHISCA} shows the probability distribution $P(\chi)$ and its scaling behavior 
for various choices of parameters. While \ref{CHISCA} (a) shows a clear dependence of the amount of non-affinity on the coarse-graining length $\Lambda$, 
figure \ref{CHISCA} (b) shows a collapse of the distributions for the scaled quantity $\chi/\Lambda^{2}$ for $\Lambda>2.2$. 
This corresponds to a minimum of $18$ neighbors to the central particle, that are taken into account in the calculation of 
the strain field via the minimization of $\chi$. These distributions show a constant offset from $\chi/\Lambda^{2}=0$. By 
contrast $\Lambda=1.3$ shows no such offset, meaning that for the calculation of the affine strain field only the  minimal 
neighborhood, i.e. the $6$ nearest neighbors, allows for a global minimization of $\chi$. In addition the probability 
distributions of the non-affine parameter $\chi$ also scale with with the spring constant $f$ (figure \ref{CHISCA} (c)) and in 
simulations run at various hydrostatic, external pressure $p$ with the resulting average density $\langle\varrho\rangle$ (figure 
\ref{CHISCA} (d)). One can therefore obtain the probability distribution for $\chi$ for any inverse temperature $\beta$, spring constant $f$, density $\rho$ and 
coarse-graining length $\Lambda$ from a generalized extreme value probability distribution function. This master curve is $P({\mathcal X}) = \left(
\left(1+0.27\cdot z\right)^{(-1/0.27)-1}e^{-\left(1+0.27\cdot z\right)^{-1/0.27}}\right)/2.012$ with $z=\left({\mathcal X}-3.127\right)/2.012$, which we 
obtain from a fit to the simulation data, with the scaling variable ${\mathcal X} = \chi \beta f/\rho \Lambda^2$, independent of system size $N$ and the choice of ensemble.

\begin{figure}
\begin{center}
\includegraphics[angle=0,width=7.25cm]{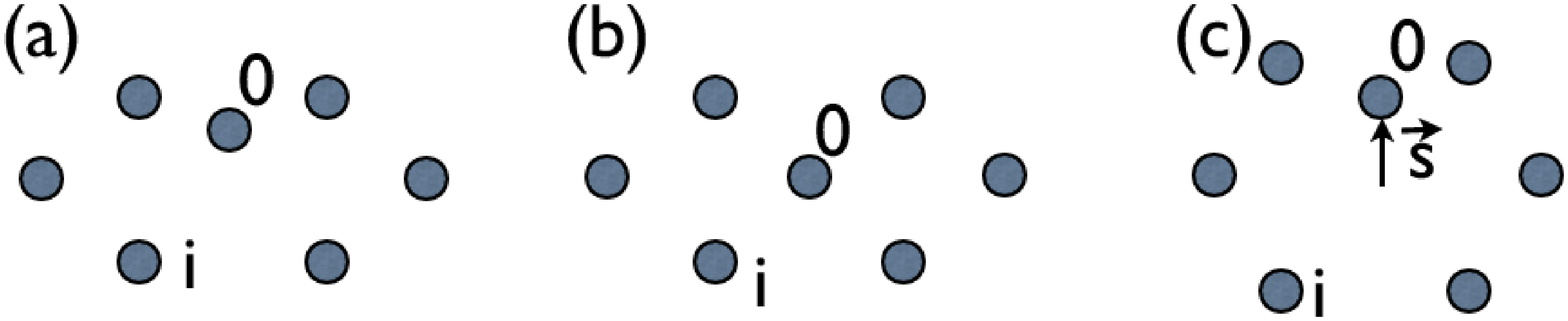}
\caption{(Color online) Example configuration (a) consisting of a central particle $0$ and neighboring particles $i = 1,6$ which can be decomposed into a purely affine deviatoric distortion (b) together with a non-affine displacement ${\vec s}$ of the central particle $0$ (c). }
\label{CAGEPIC}
\end{center}
\end{figure}

We show below that features of $P(\chi)$, like the dependence on the spring constant $f$, may be rationalized within a simple ``cell model'' calculation. In this model each 
particle is assumed  to fluctuate within the cage of its $6$ nearest neighbors which suffers, at most, an affine distortion (see figure \ref{CAGEPIC}). 
The only source of non-affinity comes from the displacement of the central particle from its equilibrium position. 
For such a subset of configurations, one may simply decompose each configuration $\{\vec{r}\}$ as that obtained by an affine 
transformation plus a non-affine displacement ${\vec s}$ of the central particle within an undistorted hexagonal cell. The 
non-affinity parameter $\chi$ may then be calculated to be, 
\begin{eqnarray}\nonumber
\chi&=&\sum_{m=1}^{6}\sum_{i=1}^{2}\big(r_{m}^{i}-r_{0}^{i}-\sum_{j=1}^{2}(\delta_{ij}+\epsilon_{ij})[R_{m}^{j}-R_{0}^{j}]\big)^{2}\\ \nonumber
&=& \sum_{m=1}^{6}\sum_{i=1}^{2}\left( r^{i}_{m} - r^{i}_{0} - (R^{i}_{m} - R^{i}_0) \right)^2 = 6 s^2  \nonumber
\end{eqnarray}
Within this approximation the ensemble average contributes to the  Lindemann parameter $l$, as $\langle|u|^{2}\rangle=\langle
(\vec{u}_{\textrm{\tiny{affine}}}+\vec{s})^2\rangle
= \langle u_{affine}^2\rangle+\langle \chi\rangle /6+2\langle |u_{\textrm{\tiny{affine}}}|*|s|\rangle =l^{2} a^{2}$. The Lindemann ratio depends on the stiffness of the solid and grows as the melting point 
is approached. Within this model the energy of these configurations can be calculated to be,
\begin{eqnarray}\nonumber
E&=&\frac{f}{2}\sum_{i}\big(\left|\Delta \vec{R}_{i}-\vec{s}\right|-\left|\Delta
\vec{R}_{i}\right|\big)^{2}=\frac{f}{2}\sum_{i}\Big\{2\Delta
\vec{R}_{i}^{2}+\vec{s}^{2}\\\nonumber
&&-2\vec{s}\cdot\Delta\vec{R}_{i}-2\Delta\vec{R}_{i}^{2}\sqrt{1+\vec{s}^{2}/\Delta\vec{R}_{i}^{2}-2\Delta\vec{R}_{i}\cdot\vec{s}/\Delta\vec{R}_{i}^{2}}~\Big\}\\\nonumber
&\approx& 3fs^{2}\nonumber
\end{eqnarray}
Here we used an approximation of the square root up to $\mathcal{O}(4)$ and the abbreviation $\Delta\vec{R}_{i}=\vec{R}_{i}-\vec{R}_{0}$. Thus the energy
related to the non-affinity $\chi$ of the central particle is $E_{cell}=E/3=f\chi/6$. With this energy contribution it is straight forward to 
calculate the probability distribution of $\chi$, 
\begin{eqnarray}\nonumber
P(\chi)&=& C\int d\vec{s}~e^{-\beta fs^{2}}\delta(\chi'(s)-\chi) \\
&=&\frac{2}{\sqrt{\pi}}\left(\frac{\beta f}{6}\right)\phi^{1/2} e^{-\phi}\label{CHIPROB}
\end{eqnarray}
where $\phi =\frac{\beta f}{6} \chi$ and $C$, the normalization constant. In figure \ref{CHISCA} (b)  we have plotted $P(\chi)/(\beta f/6)$ from this cell
model together the scaled distributions obtained from our simulations. The data collapse of the distributions from simulations with various spring constants
is in accord with the scaling in $\beta f/6$ as expected from the simple cell model. The details of the shape of the distribution function can not be
captured completely. As is to be expected in this simple model, the contributions of large $\chi$ are slightly overestimated.

\begin{figure}
\begin{center}
\includegraphics[angle=0,width=7.25cm]{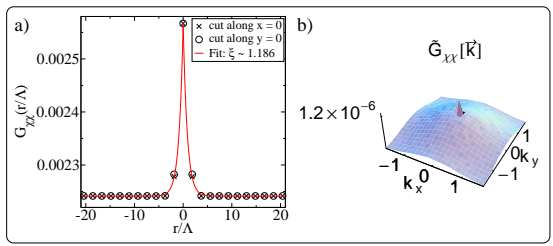}
\caption{(Color online) a) The autocorrelation function of the non-affinity parameter $\chi$ for the harmonic triangular 
lattice with spring constant $\beta a^{2} f =200/\sqrt{3}$ in the $NpT$ ensemble with periodic boundary conditions along 
the $x$ and the $y$ axes. The red line is a fit to an exponential form. b) A surface plot of its Fourier transform $\tilde{G}_{\chi\chi}(\vec{k})$.
} 
\label{CHICORR}
\end{center}
\end{figure}

How does the presence of $\chi$ influence strain correlations? To see this we assume that, at least for small $\chi$ the total strain obtained by fitting 
an arbitrary configuration to an affine transformation contains an affine part which would have been the only result if $\chi$ were zero, and a $\chi$ 
dependent non-affine part which may be expanded as a series in powers of $\chi$, namely, 
\begin{equation}\nonumber
\epsilon_{ij}(\chi) = \epsilon_{ij} + \sum_p t_p \chi^p\nonumber
\end{equation}
This decomposition is more general than what is suggested above, and it is customary, in theories of solid plasticity to decompose the
total strain into elastic (affine) and {\em plastic} (non-affine) parts $\epsilon_{ij}^T = \epsilon_{ij}^0 +
\epsilon_{ij}^P$ \cite{LUBLI}. To lowest order in $\chi$ therefore, 
\begin{equation}\nonumber
\langle \epsilon(\chi;0)_{ij} \epsilon(\chi;{\vec r})_{kl} \rangle =  \langle \epsilon^0(0)_{ij} \epsilon^0({\vec r})_{kl} \rangle + t_1^2 \langle \chi(0) \chi({\vec r}) \rangle\nonumber
\end{equation}
where we have used the fact that the coarse-graining process projects the displacements into mutually orthogonal subsets \cite{CHAIKI,MORI} so that one 
can ignore all correlations between $\epsilon_{ij}^0$ and $\chi$. In figure \ref{CHICORR} a) we have plotted cuts showing the decay of 
$G_{\chi \chi}({\vec r}) = \langle \chi(0) \chi({\vec r}) \rangle$ along the $x$- and $y$- axis. The function $G_{\chi \chi}$ is isotropic and 
decays rapidly to zero over a length scale comparable to $\Lambda$. This suggests that $\chi$ behaves as a ``delta'' correlated white noise 
with a probability distribution given by Eq. \ref{CHIPROB}. Again, this is consistent with our identification of $\chi$ with the Lindemann ratio, 
the microscopic, random, thermal fluctuations of individual particles are, indeed, expected to be uncorrelated with each other. 
Given the form of $G_{\chi \chi}$ one expects such fluctuations to contribute only a background term (compare figure \ref{CHICORR} b)) to
the strain correlations in Fourier space. 
 
We have shown in this section that the coarse-graining process by which affine strains may be extracted from microscopic particle configurations
also generates a random white noise consisting of non-affine particle displacements. For a harmonic solid this is simply related to the Lindemann
parameter, which, in turn, depends ultimately on the strength of the interactions. 

What is the general implication of this to the study of
elasticity and rheology of complex solids? The current picture of the mechanism of relaxation in amorphous materials indicates that there are two
main competing processes involved \cite{NGAI}. Over small time scales the system fluctuates within local minima in the free energy landscape 
making transition between such basins of attraction over longer time scales. We have shown here that harmonic fluctuations within local minima 
generates a well characterized contribution to the non-affine displacement $\chi$, therefore any ``extra'' contribution to $\chi$ arises 
exclusively from these inter-basin transitions. Thus our analysis may be used as a tool to distinguish between these two kinds of 
relaxations in complex solids. \\
  
\section{Conclusions}
We have shown in this paper how the analysis of particle configurations of two-dimensional soft solids gives access to a wealth of information on 
the local and non-local elastic properties. Since the harmonic solid analyzed here is the most generic conceivable our work has the potential to 
serve as a template for further research in this direction. The properties of the strain-strain correlation functions have been discussed in great
detail and various methods of how to extract the elastic moduli from their analysis were presented. Furthermore we determined and discussed the effects of external 
pressure and an embedding medium, the proper treatment of which is essential for experimentalists seeking to use our methods for analyzing mechanical behavior of 
soft matter. The implications of our work particularly for the understanding of non-affineness in solids is significant, because 
our study allows one to classify non-affine fluctuations in any system into ``trivial'' (in the sense of being present even in an ideal harmonic solid)
 and non-trivial components. In the future, we shall use these procedures to study metastability in solids undergoing phase transitions and plastic 
 behavior of solids under large external stresses. 

\begin{acknowledgments}
We acknowledge useful discussions with K. Binder, R. Messina and M. Rao. This work was funded by the Deutsche Forschungsgesellschaft (SFB TR6/C4). 
Granting of computer time from HLRS, NIC and SSP is gratefully acknowledged. One of us (SS) thanks the DST, Govt. of India for support.
\end{acknowledgments}


\begin{thebibliography}{50}
\bibitem{YETHI} A. Yethiraj, A. van Blaaderen, Nature, {\bf{421}},513 (2003).
\bibitem{HABD} P. Habdas, E. R. Weeks, Curr. Opin. Colloid Interface Sci., {\bf{7}}, 196 (2002).
\bibitem{STRAND} K.J. Strandburg, Rev. Mod. Phys., {\bf{60}}, 161 (1988); K.J Strandburg, ibid., {\bf{61}}, 747 (1989).
\bibitem{SEN_KTHNY} S. Sengupta, P. Nielaba and K. Binder, Phys. Rev. E, {\bf{61}}, 6294 (2000).
\bibitem{BIND_KTHNY} K. Binder, S. Sengupta and P. Nielaba, J. Phys.: Condens. Matter, {\bf{14}}, 2323 (2002).
\bibitem{GRUENB_MELT} H.H von Gr\"unberg, P. Keim, K. Zahn and G. Maret, Phys. Rev. Lett., {\bf{93}}, 255703 (2004).
\bibitem{WILLE} A. Wille, F. Valmont, K. Zahn and G. Maret, Europhys. Lett., {\bf{57}}, 219 (2002).
\bibitem{SEN_FLUCM} S. Sengupta, P. Nielaba, M. Rao and K. Binder, Phys. Rev. E, {\bf{61}}, 1072 (2000).
\bibitem{ZAHN} K. Zahn, A. Wille, G. Maret, S. Sengupta and P. Nielaba,  Phys. Rev. Lett., {\bf{90}}, 155506 (2003).
\bibitem{MARAG} R. Maranganti, P. Sharma, Phys. Rev. Lett., {\bf{98}}, 195504 (2007); R. Maranganti, P. Sharma, Journal of the Mechanics and Physics of Solids, {\bf{55}}, 1823 (2007).
\bibitem{KFRANZ2} K. Franzrahe, P. Keim, G. Maret, P. Nielaba and S. Sengupta, Phys. Rev. E, {\bf{78}}, 026106 (2008).
\bibitem{FALK_LA} M.L. Falk, J.S. Langer, Phys. Rev. E, {\bf{57}}, 7192 (1998).
\bibitem{LEMAI} A. Lema\^itre, Phys. Rev. Lett., {\bf{89}}, 195503 (2002).
\bibitem{MALON} C.E. Maloney, M.O. Robbins, J. Phys.: Condens. Matter, {\bf{20}}, 244128 (2008).
\bibitem{CHAIKI} P.M. Chaikin, T.C. Lubensky, \emph{Principles of condensed matter physics} (Cambridge University Press, Cambridge, UK, 1995).
\bibitem{ERINGE} D.G.B. Edelen in \emph{Continuum Physics 4}, A.C. Eringen, eds, (Academic Press,  New York, 1976).
\bibitem{GOLDF} N. Goldenfeld, \emph{Lectures on Phase Transitions and the Renormalization Group} (Westview Press, Boulder, Colorado, USA, 1992).
\bibitem{KFRANZ3} K. Franzrahe, P. Nielaba, A. Ricci, K. Binder, S. Sengupta, P. Keim and G. Maret, J. Phys.: Condens. Matter, {\bf{20}}, 404218 (2008).
\bibitem{KEIM_HARMONIC} P. Keim, G. Maret, U. Herz and H.H. von Gr\"unberg, Phys. Rev. Lett., {\bf{92}}, 215504 (2004).
\bibitem{PARR_FLUCM} M. Parrinello, A. Rahman, J. Chem. Phys., {\bf{76}}, 2662 (1982).
\bibitem{SQUIRE} D.R. Squire, A.C. Holt and W.G. Hoover, Physica, {\bf{42}}, 388 (1969).
\bibitem{WALLA} D.C. Wallace, \emph{Thermodynamics of Crystals} (Dover Publications Inc., Mineola, NY, 1998).
\bibitem{ROMAN2} F. L. Rom\'{a}n, J. A. White and S. Velasco, J. Chem. Phys., {\bf{107}}, 4635 (1997); F. L. Rom\'{a}n, J. A. White, A. Gonz\'{a}lez, S. Velasco J. Chem. Phys., {\bf{110}}, 9821 (1999).
\bibitem{LUBLI} J. Lubliner, \emph{Plasticity Theory} (Dover Publications Inc., Mineola, NY, 2008).
\bibitem{NGAI} K. L. Ngai, G. B. Wright, eds, \emph{Relaxations in Complex Systems} (NRL, Washington, DC, 1985).
\bibitem{MORI} H. Mori, Progress of Theoretical Physics, {\bf{33}}, 423 (1965).
\end{thebibliography}
\end{document}